\documentclass[fleqn,usenatbib]{mnras}

\usepackage{newtxtext,newtxmath}

\usepackage[T1]{fontenc}

\DeclareRobustCommand{\VAN}[3]{#2}
\let\VANthebibliography\thebibliography
\def\thebibliography{\DeclareRobustCommand{\VAN}[3]{##3}\VANthebibliography}

\usepackage{graphicx}	
\usepackage{amsmath}	
\usepackage{comment}
\usepackage{xcolor}

\title[21-cm power spectrum at 48 MHz from OVRO-LWA]{A 21-cm power spectrum at 48 MHz,  using the Owens Valley Long Wavelength Array}

\author[H. Garsden et al.]{
H. Garsden$^{1,2}$\thanks{E-mail: h.garsden@qmul.ac.uk (HG)},
L. Greenhill$^{1}$,
G. Bernardi$^{3,4,5}$,
A. Fialkov$^{6,7}$,
D.C. Price$^{1,8}$,
D. Mitchell$^{9}$,
J. Dowell$^{10}$,
\newauthor
M. Spinelli$^{11,12}$
and F.K. Schinzel$^{13}$
\\
$^{1}$Harvard-Smithsonian Center for Astrophysics, MS42, 60 Garden Street, Cambridge MA 02138 USA\\
$^{2}$Astronomy Unit, Queen Mary University of London, Mile End Road, London E1 4NS, United Kingdom\\
$^{3}$INAF-Istituto di Radioastronomia, via Gobetti 101, 40129, Bologna, Italy\\
$^{4}$Department of Physics \& Electronics, Artillery Road, Rhodes University, Grahamstown, South Africa\\
$^{5}$South African Radio Astronomy Observatory, FIR street, Observatory, Cape Town, South Africa\\
$^{6}$Institute of Astronomy, University of Cambridge, Madingley Road, Cambridge CB3 0HA, UK\\
$^{7}$Kavli Institute for Cosmology, Madingley Road, Cambridge, CB3 0HA, UK\\
$^{8}$International Centre for Radio Astronomy Research, Curtin University, Bentley, WA 6102, Australia\\
$^{9}$CSIRO Astronomy \& Space Science, Australia Telescope National Facility, P.O. Box 76, Epping, NSW 1710, Australia\\
$^{10}$Department of Physics and Astronomy, University of New Mexico, 210 Yale Blvd NE, Albuquerque, New Mexico\\
$^{11}$INAF-Osservatorio Astronomico di Trieste, Via G.B. Tiepolo 11, 34143 Trieste, Italy\\
$^{12}$IFPU - Institute for Fundamental Physics of the Universe, Via Beirut 2, 34014 Trieste, Italy\\
$^{13}$National Radio Astronomy Observatory, P.O. Box O, Socorro, NM 87801, USA;University of New Mexico
}

\date{Accepted XXX. Received YYY; in original form ZZZ}

\pubyear{2021}

\begin{document}
\label{firstpage}
\pagerange{\pageref{firstpage}--\pageref{lastpage}}
\maketitle

\begin{abstract}
The Large-aperture Experiment to detect the Dark Age (LEDA) was designed to measure the 21-cm signal from neutral hydrogen at Cosmic Dawn, $z \approx $15-30. 
Using observations made with the $\approx $ 200\,m diameter core of the Owens Valley Long Wavelength Array (OVRO-LWA), we present a 2-D cylindrical spatial power spectrum for data at 43.1-53.5 MHz ($z_{\rm median}\approx 28$) incoherently integrated for 4 hours, and an analysis of the array sensitivity. Power from foregrounds is localized to a ``wedge'' within $k_\perp, k_\parallel$ space. 
After calibration of visibilities using   5 bright compact sources including Vir~A, we measure $\Delta^2(k) \approx 2 \times 10^{12}\ \mathrm{mK}^2$  outside the foreground wedge, where an uncontaminated cosmological signal would lie, in principle.  The measured $\Delta^2(k)$ is an upper limit that reflects a combination of thermal instrumental and sky noise, and unmodelled systematics that scatter power from the wedge, as will be discussed. 
By differencing calibrated visibilities for close pairs of frequency channels, we suppress foreground sky structure and systematics, extract thermal noise, and use a mix of coherent and incoherent integration to simulate a noise-dominated power spectrum for a 3000\,h observation and $z= 16-37$. For suitable calibration quality, the resulting noise level, $\Delta^2(k) \approx 100$ mK$^2$ (k = 0.3 Mpc$^{-1}$), would be sufficient to detect peaks in the 21-cm spatial power spectrum due to early Ly-$\alpha$ and X-ray sources, as predicted for a range of theoretical model parameters.
\end{abstract}




\section{Introduction}

Observations of Cosmic Dawn and the Epoch of Reionization (EoR) are key to unveiling properties of the first population of stars and galaxies. The role of first galaxies in the process of reionization can be probed by the measurements of Gunn-Peterson troughs in the spectra of high-redshift quasars \citep[e.g.,][]{Fan:2006, Greig:2017, Greig:2019},  Ly-$\alpha$ emission from Lyman Break galaxies \citep{Mason:2018} and the measurement of the total optical depth to Thomson scattering of the Cosmic Microwave Background (CMB) radiation  \citep[e.g.,][]{Planck:2018}. The emergent picture shows growing evidence of late reionization ending at $z\approx 6$ \citep[e.g.,][]{Weinberger:2019}. However, these measurements  cannot 
reveal much about other processes that occur at early times, such as the heating of the Intergalactic Medium (IGM) by the first X-ray sources or the onset of primordial star formation in proto-galaxies. 

These processes can be constrained via observations of the  redshifted 21-cm line of neutral hydrogen produced at a rest-frame frequency of 1.42\, GHz \citep[e.g.,][]{2006PhR...433..181F, Barkana:2016}, which can be observed using low frequency radio telescopes today.  In addition to being sensitive to the process of reionization, the 21-cm signal depends on the temperature  of the IGM \citep[e.g.,][]{Madau:1997}, Ly-$\alpha$ radiation from stars \citep{Wouthuysen:1952, Field:1958} and the intensity of the radio background at 21 cm \citep[e.g.,][]{Feng:2018}. Therefore, this signal can be used to constrain the process of primordial star and galaxy formation \citep[e.g., probe the effect of radiative feedback mechanisms at high redshifts,  ][]{Fialkov:2013}, and probe both thermal and ionization histories of the gas.

The sky-averaged (global) 21-cm signal traces the mean evolution of the Universe and can act as a cosmic clock, revealing the timing of major cosmic events such as the onset of star formation.
The first tentative detection of the global 21-cm signal at 78\,MHz (corresponding to a redshift of $z\approx 17$) was recently reported by the EDGES collaboration \citep{2018Natur.555...67B}. This detection is not compatible with theoretical predictions based on standard astrophysical and cosmological assumptions  \citep[e.g.,][explored a large grid of 21-cm global signals, broadly varying properties of astrophysical sources within acceptable ranges]{2017MNRAS.472.1915C}, calling for alternative theories at Cosmic Dawn. The proposed solutions range from models with over-cooling of gas via dark matter interactions \citep[see][and the citation history of that work]{Barkana:2018} to the existence of extra radio background radiation in addition to the CMB \citep[e.g.,][]{2018Natur.555...67B, Feng:2018}, which is usually assumed to be the only source of  background radiation. Moreover, doubts about the signal being of cosmological origin have been expressed, with possible explanations including instrumental systematic errors  
\citep{Hills:2018, Sims:2019, Singh:2019}, ground plane interference  
\citep{Bradley:2019}, and polarized foregrounds  
\citep{Spinelli:2019}. Other existing global 21-cm experiments are attempting to verify this detection at high redshifts, including the Large-aperture Experiment to Detect the Dark Ages \citep{2016MNRAS.461.2847B, 2018MNRAS.478.4193P}, and efforts in progress with experiments such as  SARAS3, REACH\footnote{https://www.astro.phy.cam.ac.uk/research/research-projects/reach}, EDGES Mid-Band, and MIST\footnote{http://www.physics.mcgill.ca/mist/}. At lower redshifts only upper limits on the global 21-cm signal have been published so far  \citep{Monsalve:2017, Singh:2017}. 

Power spectra of  21-cm fluctuations contains information on the 
initial fluctuations in density and velocity fields, spatial distribution of sources, and their spectral properties. Power spectrum measurements can provide an important validation of the EDGES signal and aid in its theoretical interpretation \citep{Fialkov:2018, 2020MNRAS.499.5993R}.  If the EDGES signal is imprinted by astrophysical or cosmological processes at Cosmic Dawn, power spectra from $z\approx 17$ could carry a consistent signature. So far, major experimental effort has been focused on observing the EoR, with several instruments yielding upper limits in the range $z\approx 6-9$, including LOFAR \citep{LOFAR-EoR:2020},  MWA \citep{Trott:2020,Barry:2019, Li:2019}, PAPER \citep{Kolopanis:2019}, and GMRT \citep{Paciga:2013}. At the higher redshifts  corresponding to Cosmic Dawn, first attempts to measure the power spectrum have been made by \citet{EwallWice:2016} who,  using 6\,h of data from the MWA, reported a sensitivity of $10^8$\,mK$^2$ on comoving scales $k \lesssim  0.5$\,h Mpc$^{-1}$, at redshifts $11.6 \lesssim  z \lesssim 17.9$ (a few orders of magnitude above the expected 21-cm power spectrum).  \citet{2019AJ....158...84E}  used 28\,h collected by the Owens Valley Radio Observatory Long Wavelength Array \citep[OVRO-LWA,][]{2015AAS...22532801H}, and found an upper limit of $10^8$ mK$^2$ at $k \lesssim 0.10 $ Mpc$^{-1}$ at $z = 18.4$. The AARTFAAC Cosmic Explorer program \citep{2020MNRAS.499.4158G}, using the LOFAR telescope, reported an upper limit of $7.3\times 10^7$\, mK$^2$ at $k=0.144$\,h\,cMpc$^{-1}$,  $ z = 17.9-18.6 $, from 2\,h of observations.  Finally, 14 hours collected by LOFAR-LBA were used to obtain upper limits of $2.1\times 10^8$ mK$^2$  and  $2.2\times 10^8$ mK${^2}$ at $k\approx 0.038$\,h Mpc$^{-1}$ for two observed fields (3C220 and NCP)  at $z = 19.8-25.2$ \citep{Gehlot:2019}. Next-generation experiments such as HERA \citep{deboer17} and the SKA \citep{Koopmans:2015} are designed to detect the 21-cm power spectrum from a wide range of redshifts and scales, including both the EoR and Cosmic Dawn.


This paper reports  results from the  Large-aperture Experiment to detect the Dark Age (LEDA). LEDA is working towards a detection of 21-cm fluctuations in the CMB signal at Cosmic Dawn ($z \approx 15-30$), in both  global (sky-averaged) spectra and spatial power spectra, using observations from OVRO-LWA.  We are currently focused on two experiments:
1.Validation of the recent detection by \citet{2018Natur.555...67B}, and extraction of  any other 21-cm absorption/emission profiles observed at Cosmic Dawn; 2.
 Generation and analysis of power spectra of the 21-cm signal at Cosmic Dawn, and analysis of the sensitivity of power spectra as a function of observation integration time.

A paper reporting on progress on the first experiment is under review \citep{2020arXiv201103994S}; this paper reports results from the second experiment.

The first power spectrum obtained from OVRO-LWA observations \citep[][mentioned above]{2019AJ....158...84E} was an angular power spectrum generated from 28\,h of integrated data using    m-mode analysis \citep{Shaw:2014},  and filtering of foregrounds using  the double Karhunen-Lo\'eve transform. 
In this paper we also use observations from OVRO-LWA, but   generate power spectra using the ``delay spectrum'' method \citep{2012ApJ...756..165P}, which produces a 2-D cylindrical power spectrum. Instead of removing foregrounds, the delay spectrum method isolates foreground power in a ``wedge'' region of the power spectrum, outside of which the 21-cm signal may be detected, given enough power spectrum sensitivity. Foreground isolation is inherent in the delay spectrum method, and relies on the spectral smoothness of foreground emission.  The method also has the advantage that it involves minimal manipulation of   telescope data, and is easy to implement. It does, however, have limitations, which we will discuss below. We  operate at a higher redshift of 
$z \approx 28$, compared to \cite{2019AJ....158...84E}, producing a power spectrum at a redshift higher than any other reported to date. On the other hand, we use a smaller number of integrated observations than \cite{2019AJ....158...84E}, namely 4\,h.

We will show that the sensitivity of our power spectrum is low compared to other reported experiments, but that this can be improved by integrating a larger number of observations, and varying the method of integration. If some or all observations are coherently integrated before power spectrum generation, the noise in the power spectrum is substantially reduced. However the timing of observations is crucial for coherent integration, since observations must be nearly identical, apart from noise. We report the results of simulations and  analysis showing that 21-cm Cosmic Dawn fluctuations can be detected using OVRO-LWA observations that are combined   using a mix of coherent and incoherent integration,  and if the observations and power spectrum are free of any systematics. The number of observations required to detect 21-cm fluctuations is large, i.e. 3000\,h in total, but it can be obtained within a reasonable time frame using a scheme that we will describe.

The paper is organized as follows: Section \ref{ovro_dec} describes the Owens Valley Long Wavelength Array (OVRO-LWA) and data processing; Section \ref{data_description} describes the observations used for power spectrum generation, and how they were selected; Section \ref{get_power} presents the delay spectrum method; Section \ref{results} shows the power spectra generated from observations, and discusses their characteristics; Section \ref{sensitivity_analysis}  describes simulations of noise in power spectra, introduces coherent integration, and reports analysis of the sensitivity of OVRO-LWA;
Section \ref{cosmic_dawn_with_ovro} compares that sensitivity   against theoretical models of Cosmic Dawn and demonstrates that OVRO-LWA, and thus compact interferometers, could be used to study Cosmic Dawn;  Section \ref{summary} summarizes the results and future work.

For this work we assume a flat universe with the following cosmological parameter values: $\Omega_\mathrm{M} = 0.31, \Omega_\mathrm{\Lambda} = 0.69, H_\mathrm{0} = 0.68$.


\section{The OVRO-LWA Telescope and Processing Pipeline}
\label{ovro_dec}
\begin{figure}
\includegraphics[width=8cm]{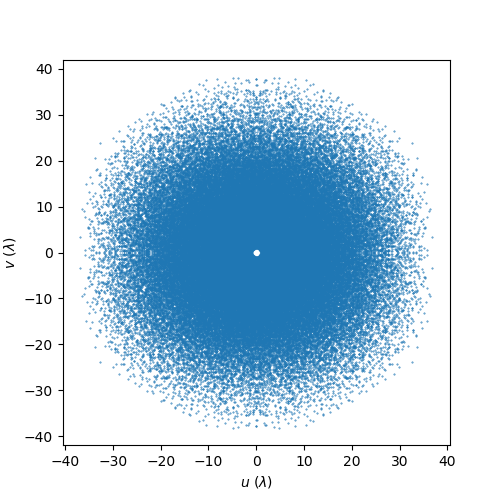}
\caption{$uv$-coverage of the  OVRO-LWA baselines used for power spectrum generation,  obtained from a frequency channel of width 24 kHz at a frequency of 57 MHz.}
\label{uvcov}
\end{figure}

OVRO-LWA is a wide field, low frequency, drift-scan, non-redundant, imaging interferometer situated in Owens Valley, California, at a latitude of 
$37.2398^\circ$ and a longitude of $-118.2817^\circ$. Its design and hardware components are based on that of the Long Wavelength Array (LWA) in New Mexico \citep{2012JAI.....150004T}. OVRO-LWA provides data for research in several  areas of  observational astronomy and cosmology, including: cosmic rays \citep{ 2019ICRC...36..405R,2019ICRC...36..211C,2020NIMPA.95363086M}, radio transients surveys \citep{2019ApJ...886..123A}, solar weather and coronal mass ejections \citep{ 2017reph.conf40101H,2019AGUFMSH21B..03C}, exoplanet searches \citep{2017reph.conf40102A}, radio counterparts to gamma ray bursts  and gravitational waves \citep{2018ApJ...864...22A,2019GCN.24196....1A}, emission from compact binaries \citep{2019APS..APRQ16006C}, properties of the ionosphere \citep{2017AGUFMSA21A2499S}, sky surveys at low frequencies \citep{2018IAUS..333..110E}, Cosmic Dawn  21-cm global signal detection \citep{2018MNRAS.478.4193P}, and Cosmic Dawn 21-cm power spectra \citep{2019AJ....158...84E}.

The telescope  consists of 256 dual-antenna stands, of the same type used in the Long Wavelength Array \citep{2012JAI.....150004T}; only the OVRO-LWA core, consisting of 219 stands, was used for  experiments reported here. The  219 stands provide 23871 baselines, with lengths from 4.8 to 212.4 meters. The observing bandwidth is 27.384-84.912 MHz (z=15.7 to 50.9 for the 21-cm HI line). The $uv$-coverage  of the OVRO-LWA core at 57\,MHz  is shown in Figure \ref{uvcov}; $uv$ lengths range from 0.9 to 40.4 wavelengths.  The resolution of the telescope is  $1.4^\circ$ at 57\,MHz.

Signals from the antennas are processed by an FX correlator on site at the Owens Valley Radio Observatory \citep{2015JAI.....450003K}. The observing band is split into 22 sub-bands each containing 109 channels of width 24\, kHz, 2398 channels in all. Each stand has a folded cross-dipole, for observing  X (east-west) and Y (north-south) polarizations. Cross-correlation of antennas generates four polarization products: XX, XY, YX, YY.   Only  XX and YY polarizations are used for our power spectrum analysis, and they are treated separately.  

Antenna signals are correlated, integrated, and saved as visibilities every 9\,s, thus producing a single ``observation'' every 9\,s.  
Multiple observations are typically  integrated for further processing, e.g. for generating power spectra.   Observations are calibrated using the Real Time System  (RTS), which performs self-calibration against a point source sky model \citep{2008ISTSP...2..707M}. The RTS uses a beam model  described in LWA Memo 178\footnote{http://www.phys.unm.edu/~lwa/memos/memo/lwa0178a.pdf}. RTS contains an algorithm to detect and flag narrow-band RFI in each sub-band,  by finding channels whose amplitude varies significantly  from the  median amplitude  of the sub-band \citep{2010rfim.workE..16M}.  
No other automated flagging is applied; further flagging of data is achieved by a visual inspection of antenna autocorrelations. Post-calibration quality checks (described below), are used to select observations to be used for science experiments.

\begin{figure*}
    \begin{minipage}[t]{\columnwidth}
        \centering
        \includegraphics[width=\columnwidth]{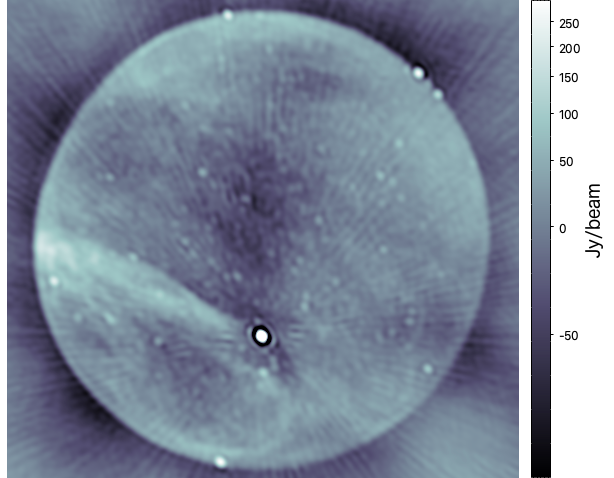}
    \end{minipage}
    \hspace{0.2in}
    \begin{minipage}[t]{\columnwidth}
        \centering
        \includegraphics[width=\columnwidth]{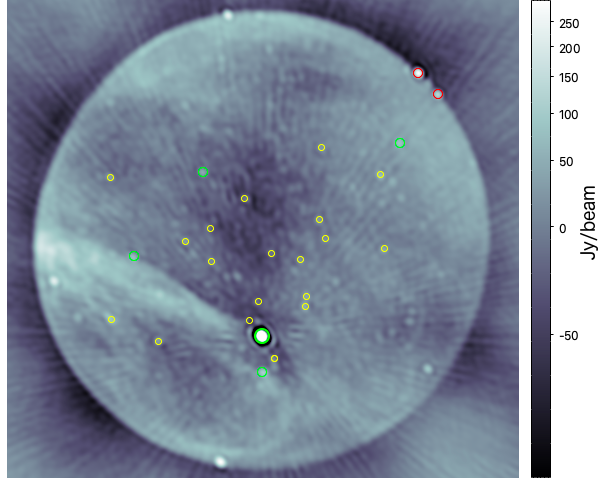}
    \end{minipage}
    \caption{An image of the sky generated from an  OVRO-LWA observation used for power spectrum generation, at LST = 12:30.  The right image is the same as the left, but is marked with sources of interest. The 5 brightest sources (after primary beam correction) at this LST were used for calibration. They are ringed in green, and include Virgo A, the brightest source, which is at zenith. The   20  next brightest sources, after the 5 used for calibration,   are ringed in yellow. The two sources on the horizon, ringed in red, are intermittent RFI sources; they are possibly located  in the towns of Bishop or Mammoth Lakes, both north-west of OVRO-LWA at distances of 18 and 76\,km respectively. The source on the horizon at the top is Cassiopeia A, the source near the horizon at the bottom is unidentified. The images contain Stokes I observed over a bandwidth of 10.464\,MHz centered at 48.324\,MHz. They were generated using NRAO CASA with Briggs weighting applied (robust=0.5). To enhance the visibility of the dimmer sources, an Arcsinh color scale is used, and  the image is clipped to a range of -90 to 300 Jy/beam. The image has not been CLEANed. The confusion limit for OVRO-LWA at this frequency, based on the longest baseline length of 200\,m, is 96\,Jy/beam.   }  
    \label{the_sky}
\end{figure*}

 \begin{table}
\centering
    \begin{tabular}{|c|c|c|c|}
    \hline
    Source  & Location & Flux Density at & Flux Density at\\
        &  (RA/DEC)  &   74 MHz (Jy) & 48.324 MHz (Jy)\\
    \hline\hline
     Virgo A  &  12:30:49  & 1253.4    &  1689.03\\
              &   +12:23:28 & &\\
     \hline
     1411+5212  & 14:11:20  & 120.27  & 162.07\\
                &+52:12:04 & &\\
     \hline
     1229+0202 &  12:29:06  & 149.96  &  202.08 \\
               &   +02:02:56 & &\\
     \hline
     1504+2600  &  15:04:59  & 129.86 &  174.99\\
                &   +26:00:46 & &\\
     \hline
     0813+4813  &  08:13:36.34  & 136.40 & 183.80\\
                &  +48:13:01 & &\\
     \hline
    \end{tabular}
    \caption{Details of the five sources used for calibration of the OVRO-LWA observations used for power spectrum generation. Locations and flux densities were obtained from the VLA Low-Frequency Sky Survey at 74\,MHz, and flux densities converted by RTS to  OVRO-LWA  frequencies by assuming a spectral index of -0.7. The flux densities at 74\,MHz and 48.324\,MHz are shown.}
    \label{sky_sources}
\end{table}

\section{Observations used for Power Spectrum Generation}
\label{data_description}

We use 4 hours  of non-contiguous (in time)  observations, selected from all observations made during May 2018, between LST 11:55 and 13:15.
Of the 219 stands available for power spectrum observations, only 165 were deemed usable over the entire 4 hours, the rest being flagged due to hardware issues and outages. We also flagged 509 baselines due to suspected cross-talk between their antenna cable connections at backend electronic processing units\footnote{M. Eastwood, personal communication.}. During calibration, short baselines were downweighted so that any observed diffuse emission does not interfere with  calibration by RTS, which lacks a diffuse emission model.

During observations, both the Sun and Galactic Center were below the horizon, and two strong sources, Cygnus A and Cassiopeia A, were on or below the horizon. The brightest source in the sky was Virgo A, at an elevation of $\approx 63^\circ$.  We used Virgo A and four other sources   for calibration.
These five sources were selected based on the following criteria:
1. at an LST of 12:30, they are above an elevation of $30^\circ$, 2. at an LST of 12:30, they are determined by RTS to be the five brightest sources after primary beam correction. The LST of 12:30 in these criteria  was chosen as it is the approximate middle of the LST range of 11:55 to 13:15.    All observations were separately calibrated using the same five sources. The RTS was configured to use the VLA Low-Frequency Sky Survey at 74\,MHz  \citep{2007AJ....134.1245C} to determine source flux densities and locations;  the sources are detailed in Table \ref{sky_sources}. To obtain flux density at frequencies other than 74 MHz we assume a spectral index of -0.7 for all sources. It is possible that the spectrum of these sources turns over at lower frequencies, but only $\approx 4.5\%$ of sources in the GLEAM catalog do so \citep{Callingham}, and Virgo A does not \citep{2012A&A...547A..56D}, so we do not attempt to correct for this.

A simple method was used to select the 4\,h of observations used for power spectrum generation, based on the  quality of the observations and their suitability for use in power spectra. The method is described  in Appendix \ref{selection_snapshots}. This produced a set of observations that are not necessarily contiguous in time, and  recorded on different days.
Figure \ref{the_sky} shows  Stokes I images of the sky generated from one observation, at LST = 12:30.  The sky is the same in both images, but the right image marks the calibration sources in green, and other sources of interest in red and yellow (see caption). The images were obtained using NRAO CASA software\footnote{https://casa.nrao.edu}.

For  power spectra we  use a frequency  band of 10.464\,MHz centered at 48.324\,MHz. That band was chosen because it is less subject to RFI and produces the highest signal-to-noise ratio over the OVRO-LWA band. It does not cover the EDGES detection frequency, but for our sensitivity analysis, using simulations, we will do so by expanding the frequency range to cover almost the entire OVRO-LWA band. Over the 10.464\,MHz band, the 21-cm signal is observed at $z = 25-31$. This corresponds to a time range of 35 million years, when the Universe was 100\,Myr old. Power spectra of 21-cm fluctuations should ideally be obtained over a time range when the Universe is not significantly evolving, which is unlikely over $z = 25-31$, so a narrower frequency band is more appropriate. However, that will produce a power spectrum with a less resolution, and  we prefer to maintain a higher resolution for  analysis of our first power spectrum  (see \cite{2013ApJ...768L..36P} for a similar argument).

\begin{figure*}
    \begin{minipage}[t]{\columnwidth}
        \centering
        \includegraphics[width=\columnwidth]{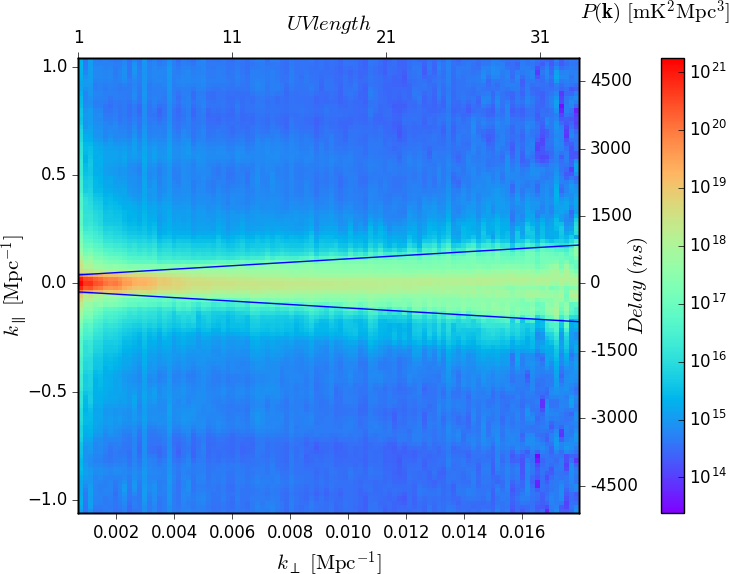}
       
    \end{minipage}
    \hspace{0.2in}
    \begin{minipage}[t]{\columnwidth}
        \centering
        \includegraphics[width=\columnwidth]{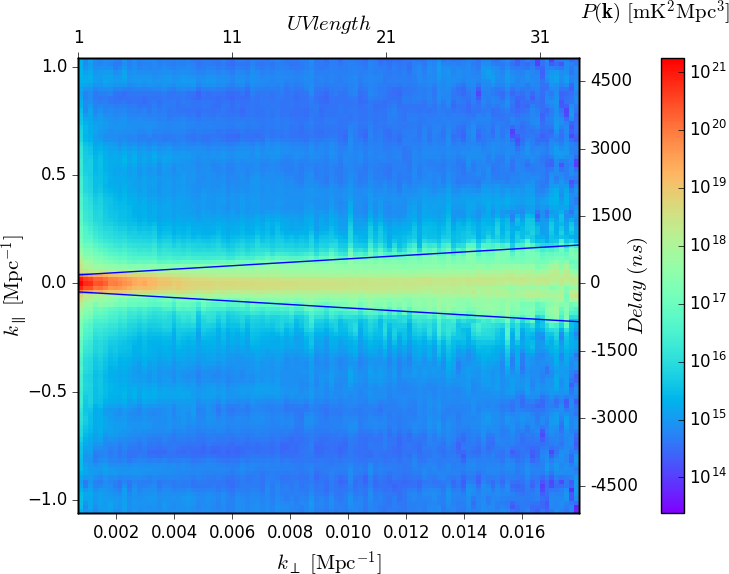}
       
          \vspace{0.15in}
    \end{minipage}

    \begin{minipage}[t]{\columnwidth}
        \centering
        \includegraphics[width=\columnwidth]{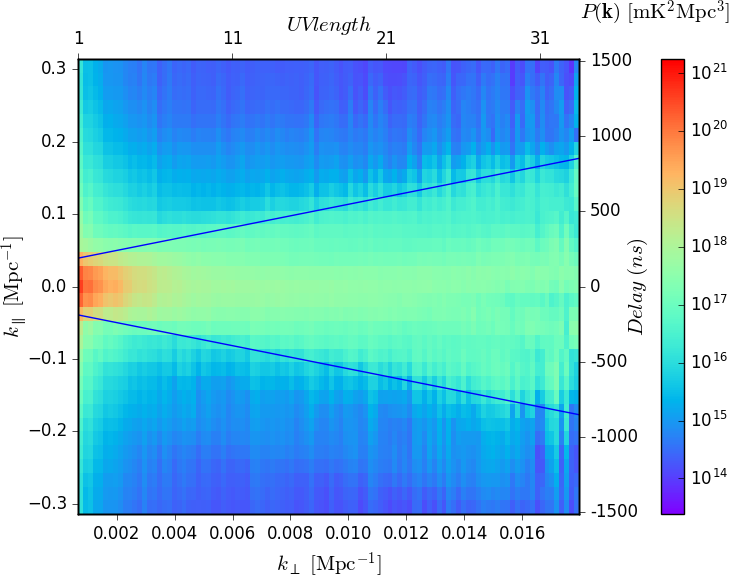}
      
    \end{minipage}
    \hspace{0.2in}
    \begin{minipage}[t]{\columnwidth}
        \centering
        \includegraphics[width=\columnwidth]{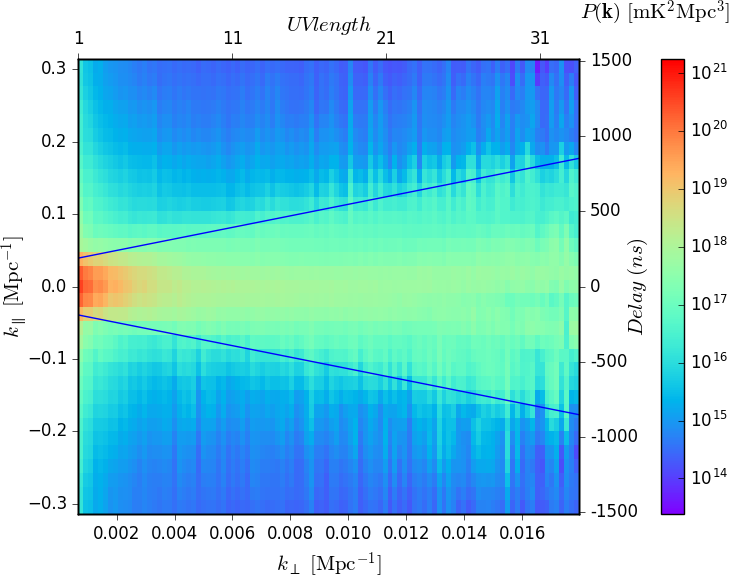}
          \vspace{0.2in}
    \end{minipage}
         
       \begin{minipage}[t]{\columnwidth}
        \centering
        \includegraphics[width=0.9\columnwidth]{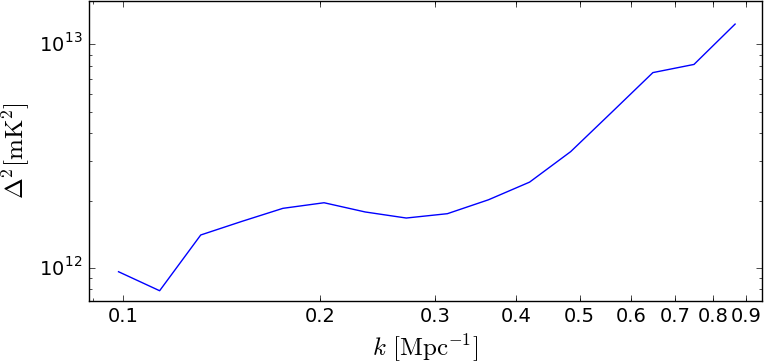}\\
       
    \end{minipage}
            \hspace{0.2in}
    \begin{minipage}[t]{\columnwidth}
        \centering
        \includegraphics[width=0.9\columnwidth]{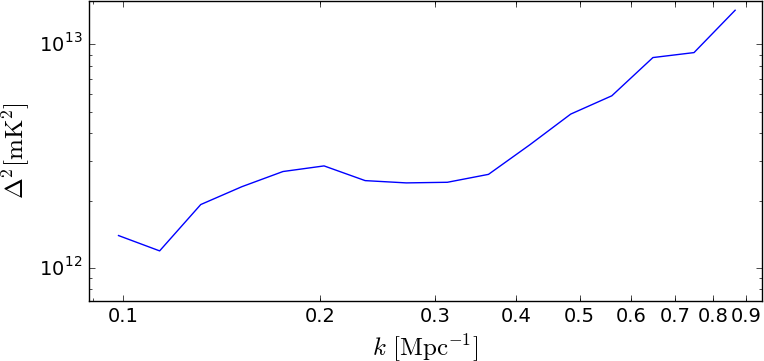}\\

    \end{minipage}

    \caption{The power spectra generated from the  4\,h of observations  described in the text, and plots of  $\Delta^2(k)$ obtained from the power spectra. The top row shows the power spectra,  the second row shows the same power spectra but restricts the $k_\parallel$ range to $-0.31$ to $0.31$\,Mpc$^{-1}$, so that the wedge can be seen in more detail. Power values are colored using a log scale. The third row plots $\Delta^2(k)$ values at various $k$, obtained from within the Cosmic Dawn window of the power spectra (the exact region used is described in Appendix  \ref{appendix_regions}). Some important properties of the power spectrum images in the top row are listed in Table \ref{properties}. The left column shows the power spectrum generated from polarization XX, the right column from polarization YY.}
    \label{power_spectrum}
\end{figure*}

\begin{table}
\centering
    \begin{tabular}{|c|c|}
    \hline
    Property  & Value\\
    \hline\hline
     Power spectrum  dimensions (pixels)&  102 horizontal, 108 vertical\\
     \hline
     $k_\perp$ range & 0.001  to 0.018 Mpc${^{-1}}$\\
     \hline
     $k_\perp$ resolution & 0.000175 Mpc${^{-1}}$\\
     \hline
     $k_\parallel$ range &  -1.06 to 1.04 Mpc${^{-1}}$\\
     \hline
     $k_\parallel$ resolution &   0.02 Mpc${^{-1}}$\\
     \hline
     Delay range & -5113.64 to 5018.94 ns \\
     \hline
     Delay resolution & 95 ns\\
     \hline
     Horizon delay at max $k_\perp$ & 852 ns\\
     \hline
     $uv$ lengths used & 1 to 33\\
     \hline
    \end{tabular}
    \caption{Size, ranges, and resolution of the power spectrum images in the top row of Figure \ref{power_spectrum}.}
    \label{properties}
\end{table}

\begin{table}
\centering
\begin{tabular}{|c|c|c|}
\hline
              &  Pol XX  & Pol YY   \\
              \hline\hline
 Observed $P(\mathbf{k})$, foreground wedge &    $1.1 \times 10^{18}$      &      $1.7 \times 10^{18}$    \\
 \hline
 Observed $P(\mathbf{k})$, Cosmic Dawn window &    $5.2 \times 10^{14}$    &  $6.7 \times 10^{14}$   \\
 \hline
 Ratio of $P(\mathbf{k})$ above                  & 2039    & 2490   \\
 \hline
 $\Delta^2(k = 0.3)$, Cosmic Dawn window                & $1.7 \times 10^{12}$    & $2.4 \times 10^{12}$  \\
 \hline
\end{tabular}
\caption{The top two rows show the average power in the power spectra generated from 4\,h of observations (Figure \ref{power_spectrum}), from the foreground wedge (first row), and the Cosmic Dawn window (second row). The third row shows the ratio of the power in the foreground wedge to the power in the Cosmic Dawn window. The bottom row shows values of dimensionless power $\Delta^2(k = 0.3)$ obtained from the Cosmic Dawn window. }
\label{psvalue}
\end{table}

\begin{figure*}
    \begin{minipage}[t]{\columnwidth}
        \centering
        \includegraphics[width=\columnwidth]{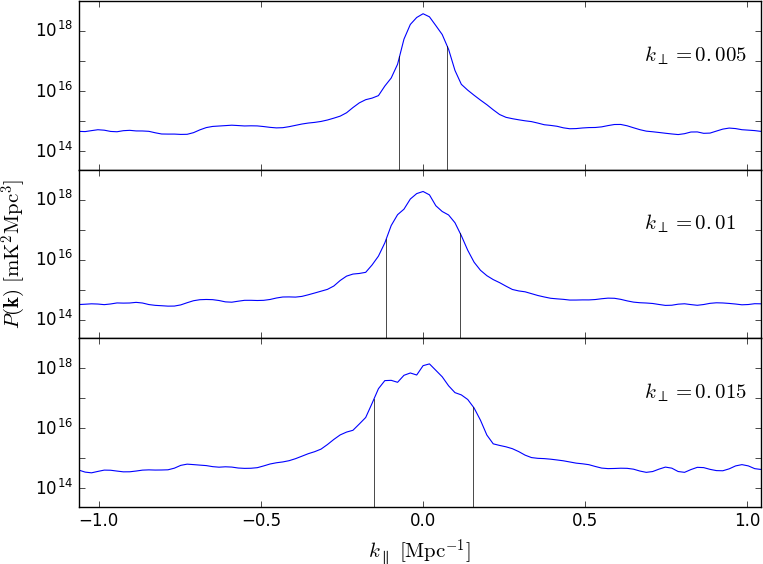}
         
    \end{minipage}
    \hspace{0.2in}
    \begin{minipage}[t]{\columnwidth}
        \centering
        \includegraphics[width=\columnwidth]{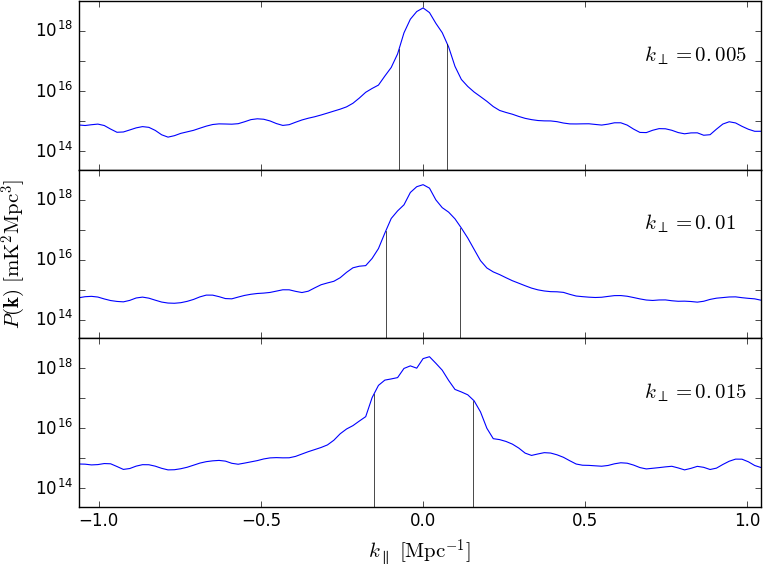}
        
            \vspace{0.15in}
    \end{minipage}

    \begin{minipage}[t]{\columnwidth}
        \centering
        \includegraphics[width=\columnwidth]{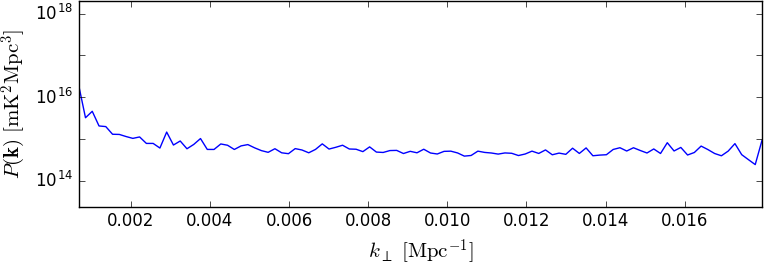}
        
    \end{minipage}
    \hspace{0.2in}
    \begin{minipage}[t]{\columnwidth}
        \centering
        \includegraphics[width=\columnwidth]{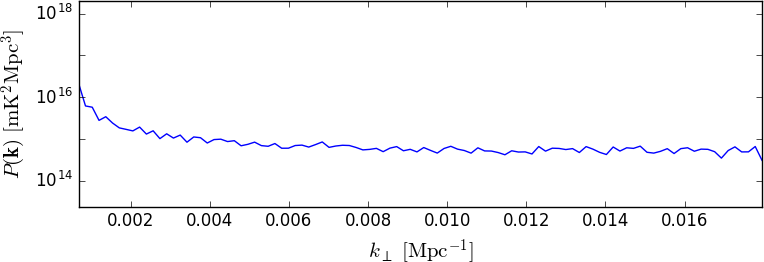}
         
         \vspace{.1in}
    \end{minipage}
    \caption{Vertical and horizontal cuts through the power spectra in the top row of Figure \ref{power_spectrum}. The top row panels contain  3 vertical cuts  made at
      $k_\perp$ values of 0.005, 0.01, 0.15\,Mpc$^{-1}$. The black lines
    indicate the location of the horizon at these $k_\perp$ values.  The bottom row shows a horizontal cut made at $k_\parallel = 0.53\, \mathrm{Mpc}^{-1}$. The left column is from power spectra generated from polarization XX, the right column from polarization YY.}
    \label{cut_ps}
\end{figure*}

\section{Power Spectrum Generation}
\label{get_power}
OVRO-LWA observes   21-cm fluctuations at redshifted 21-cm frequencies,  as well as foregrounds emitting at the same  frequencies; foregrounds signals are dominated by   Galactic synchrotron and free-free emission, and extragalactic sources. To separate the foregrounds from the 21-cm signal in power spectra, several methods have been used \citep{2019MNRAS.483.2207M}, which fall broadly into two categories: foreground removal, and foreground isolation.  Examples of the first are reported in \citet{Gehlot:2019}, \citet{2019AJ....158...84E} and \citet{Barry:2019}; examples of the second in   \citet{2013ApJ...768L..36P},   \citet{2015ApJ...804...14T}, \citet{2016ApJ...833..213P} and \citet{EwallWice:2016}. We implement foreground isolation by  generating  a 2-D,  cylindrically averaged, power spectrum using the delay spectrum method \citet{2012ApJ...756..165P}.

The delay spectrum method applies a Fourier transform  to baselines visibilities, producing a delay spectrum:  
\begin{equation}
    \mathcal{V}_\mathrm{b}(\tau) = \int V_\mathrm{b}(\nu) e^{-2 \pi \nu \tau } d\nu,
    \label{delayft}
\end{equation}
where  $V_\mathrm{b}(\nu)$ are visibilities for a baseline (arrayed by frequency), and $\tau$ is the Fourier conjugate of $\nu$. $\tau$ represents the time delay of foreground signals between the two antennas of the baseline, and foregrounds signals are confined to values of $\tau$ less than the horizon delay, thus producing the ``wedge''. The baseline visibilities $V_\mathrm{b}(\nu)$ measure the spatial Fourier transform perpendicular to the line of sight, and can all be assigned a  wavenumber $k_\perp$. Equation \ref{delayft} also approximates the spatial Fourier transform of the 21-cm signal along the line-of-sight, with a wavenumber $k_\parallel$ assigned to $\tau$. 
 
 The delay spectra from multiple baselines are converted to power and binned by $k_\perp$,  producing  a 2-D power spectrum indexed by ($k_\perp, k_\parallel)$.
Power is calculated using the equation
\begin{equation}
P(\mathbf{k}) \approx |\mathcal{V}|^2 \left(\frac{{\lambda}^2}{2 k_\mathrm{B}}\right)^2 \frac{X^2 Y}{\Omega_\mathrm{pp} B_\mathrm{pp}}
\label{P_eqn}
\end{equation}
\citep{2012ApJ...753...81P, 2014ApJ...788..106P, ParsonsMemo},
where  $P(\mathbf{k}$) is the power at $\mathbf{k}$ = ($k_\perp, k_\parallel), \lambda$ is the mean wavelength of the band, $k_\mathrm{B}$ is Boltzmann's constant,  
$X^2 Y$ converts angles and frequency intervals to comoving distance \citep{1999astro.ph..5116H}, $\Omega_\mathrm{pp}$ is the integral over the power-squared beam \footnote{Power-squared indicated by the double p subscript}, and $B_\mathrm{pp}$ is the effective bandwidth. $B_\mathrm{pp} = \int |w|^2 d\nu$, where $w$ is the window function applied to the delay transform. A window function is used to avoid spectral leakage produced by taking the Fourier transform of a finite length of non-periodic data; we use a Blackman-Harris window. We use a beam model of the LWA antenna dipole \citep{DowellMemo} to obtain $\Omega_\mathrm{pp}$. The units of $P(\mathbf{k})$ are $\mathrm{ mK}^2 \mathrm{Mpc}^3$. $P(\mathbf{k})$ is often converted to  dimensionless power 
\begin{equation}
    \Delta^2(k) = \frac{k^3}{2 \pi^2} P(\mathbf{k}),
    \label{delta_eqn}
\end{equation}
where $k = \sqrt{k_\perp^2 + k_\parallel^2}$, and $\Delta^2(k)$ has units of $\mathrm{mK}^2$.
We will use both power values in this paper.

The  delay spectrum method  has limitations due its approximation of the spatial Fourier transform along the line-of-sight.    The delay spectrum method works best for narrow-beam, short-baseline, interferometers \citep{2012ApJ...756..165P}, whereas OVRO-LWA is a wide-field, long-baseline interferometer. \citet[Eqn. 9]{2014PhRvD..90b3018L} have determined that the delay-spectrum method may be used for observations where 
\begin{equation}
b \theta_\mathrm{0} \ll \frac{c}{B_\mathrm{band}},
\label{appr_limit}
\end{equation}
where $b$ is the length of a baseline, $\theta_\mathrm{0}$ is the width of the telescope primary beam (1.8 radians), 
and $B_\mathrm{band}$ is the bandwidth over which the observations are made.  Interpreting ``$\ll$'' as ``one-half'', Equation \ref{appr_limit} is violated by 99.66\% of the OVRO-LWA baselines used for our power spectrum generation, indicating that the baselines are too long and/or the bandwidth too wide. Reducing the maximum baseline length will reduce the  $k_\perp$ range;  reducing the bandwidth will reduce the $k_\parallel$ range. For the power spectra reported in this work, we ignore these restrictions, so as to obtain a power spectrum with an extent that allows for some analysis. We will deal with them in future work (see Section \ref{summary}).

\subsection{Additional processing steps}
\label{additional_steps}

Before generating a delay spectrum from each baseline, flagged channels must be dealt with. Flagged channels represent missing values in the array $V_\mathrm{b}(\nu)$ (Equation \ref{delayft}), which produce a delay spectrum convolved with a PSF. This PSF may be deconvolved using a CLEAN algorithm \citep{1974A&AS...15..417H}. We use an implementation of CLEAN developed by the HERA Consortium\footnote{\texttt{pspec\_prep.py}, https://github.com/HERA-Team} for just this purpose. CLEAN is applied to the delay spectrum within the horizon, including a buffer of 95\,ns  outside the horizon; 95\,ns being the resolution of $\tau$  in the delay spectrum of OVRO-LWA baselines (Table \ref{properties}).

After  CLEAN, the visibilities in $V_\mathrm{b}(\nu)$ are averaged by groups of 4, to reduce noise.  This reduces the frequency resolution by a factor of 4. The bandwidth of 10.464\,MHz, which initially contains 436 channels of width 24 kHz, is thus transformed to 109 channels  of width 96\, kHz.

Instead of converting each baseline delay-spectrum to power, 
we follow \cite{2013ApJ...768L..36P} and multiply observing-time adjacent (i.e. LST adjacent) pairs of the same baseline, to reduce noise. The LSTs of each multiplied pair must be within the time smearing limits for OVRO-LWA (see Appendix \ref{integration_time_limit}), since they are being coherently combined via the multiplication. Fortunately, after ordering all observations by LST, this applies to 99.9\% of the observations comprising the 4\,h. 

\section{Results}
\label{results}
Figure \ref{power_spectrum} shows the power spectrum generated from the 4\,h  of observations described in Section \ref{data_description}, for  polarizations XX and YY.  Baselines are binned, i.e.averaged, along the $k_\perp$ axis, using a bin size of 0.000175\,Mpc$^{-1}$, being the $k_\perp$ of a baseline with u $= 0.5$ at the lowest observing frequency. There is no binning along the $k_\parallel$ axis. The blue lines are the horizon delay, which increases with increasing $k_\perp$, producing the ``wedge'' within which foreground emission is isolated. $k_\parallel$ wavenumbers are approximated from the delays $\tau$ of the delay transform; these delays are plotted on the y-axis on the right.  $k_\perp$ values approximately correlate with baseline $uv$ lengths, which are shown on the top x-axis.  

The top row of Figure \ref{power_spectrum} displays the power spectra using the complete range of $k_\parallel$ values derived from the observing frequency band (10.464\,MHz); the second row is the same power spectrum but restricts the $k_\parallel$ range to $\pm 0.3$\,Mpc$^{-1}$ so that the wedge can be seen in more detail. A summary of the important properties of the power spectrum images in the top row of Figure \ref{power_spectrum} is in Table \ref{properties}. The third row of Figure \ref{power_spectrum} shows the dimensionless power  $\Delta^2(k)$  extracted from the Cosmic Dawn window, at various values of $k$ (the algorithm  for extracting $\Delta^2(k)$ is described in Appendix \ref{appendix_regions}).
\begin{figure}
    \centering
    \includegraphics[width=\columnwidth]{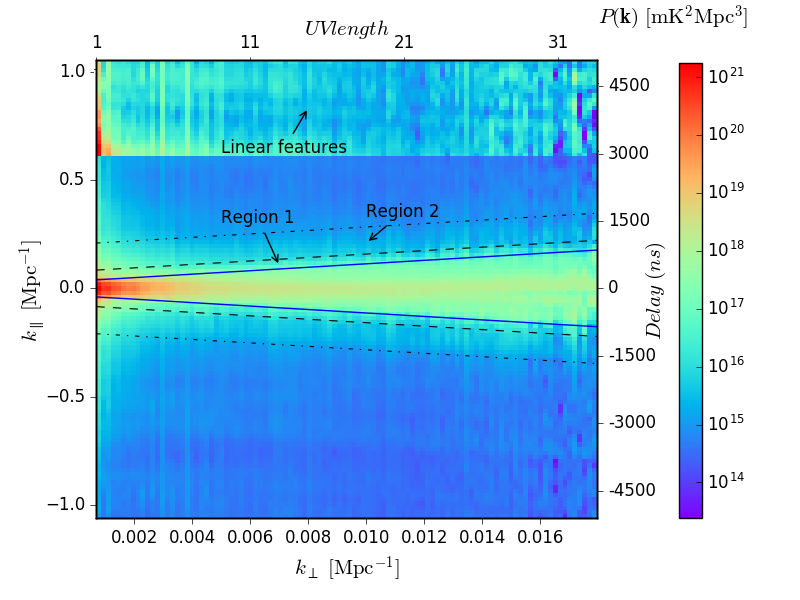}
    \caption{The power spectrum for XX polarization, (Figure \ref{power_spectrum}, top left image) overlaid with features of 
    interest. Regions 1 and 2 are regions of spillover in the Cosmic Dawn window. The top section  indicates linear features that  are approximately horizontal, with a slight downward slope, and run across almost the entire $k_\perp$ range. The top section  has been clipped to $10^{14} <P(\mathbf{k}) < 10^{16}$ mK$^2$Mpc$^3$, and rescaled to make the features clearer.}
    \label{plateau}
\end{figure}

Figure \ref{cut_ps} shows vertical and horizontal cuts through the power spectra. The top row depicts vertical cuts  at three  $k_\perp$ values of 0.005, 0.01, 0.015\,Mpc$^{-1}$. The location of the horizon at the different $k_\perp$ values is shown  by the black vertical lines. The bottom row of Figure \ref{cut_ps} depicts a horizontal cut through the power spectra at a constant $k_\parallel = 0.53$\,Mpc$^{-1}$. These cuts show that the power drops by a factor of 10$^4$\,mK$^{2}$Mpc$^3$ as $k_\parallel$ moves away from 0, and that the power in the Cosmic Dawn window is fairly flat by $k_\perp$. The average power levels $P(\mathbf{k})$ and $\Delta^2(k)$ within  the foreground wedge and Cosmic Dawn window are listed in Table \ref{psvalue}.

\begin{figure*}
    \begin{minipage}[t]{\columnwidth}
        \centering
        \includegraphics[width=\columnwidth]{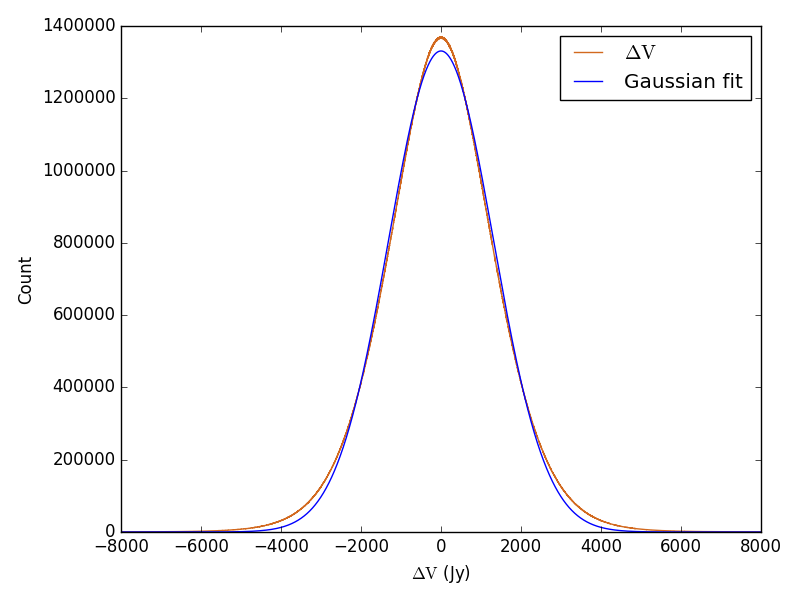}
      
    \end{minipage}
    \begin{minipage}[t]{\columnwidth}
        \centering
        \includegraphics[width=\columnwidth]{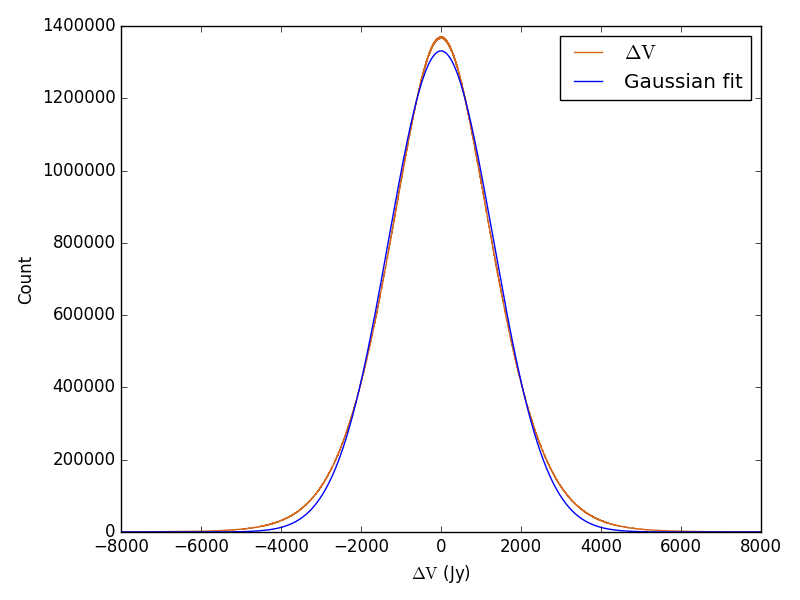}
  
            \vspace{0.05in}
    \end{minipage}
    \\

        \begin{minipage}[t]{\columnwidth}
        \centering
        \includegraphics[width=\columnwidth]{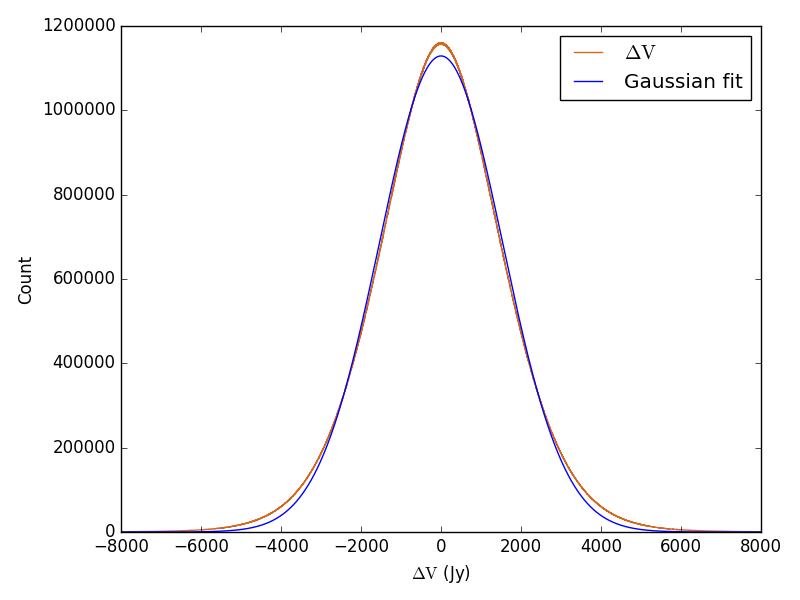}
     
    \end{minipage}
    \begin{minipage}[t]{\columnwidth}
        \centering
        \includegraphics[width=\columnwidth]{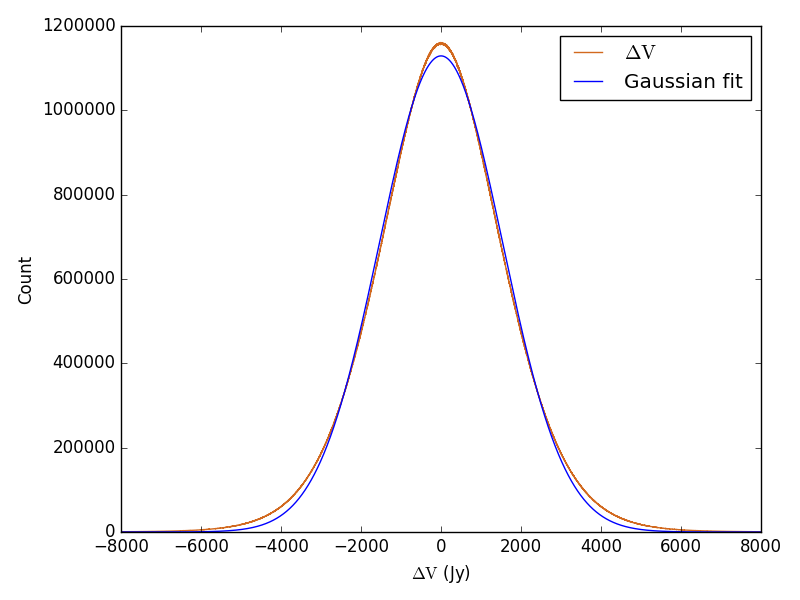}
        
    \end{minipage}

    \caption{Histograms of noise values ($\Delta V$) obtained from differencing visibilities in adjacent frequency channels, over all observations used to generate power spectra. These represent the distribution of thermal noise in OVRO-LWA (although some systematics  are present in the noise distributions, as described in the text). The top row is obtained from the XX polarization visibilities, the bottom row is from polarization YY. The real and imaginary values of the complex
    noise values are shown separately; the real values in the left column, and imaginary values in the right.}
    \label{noise_values}
\end{figure*}


\begin{figure*}
    \begin{minipage}[t]{\columnwidth}
        \centering
        \includegraphics[width=\columnwidth]{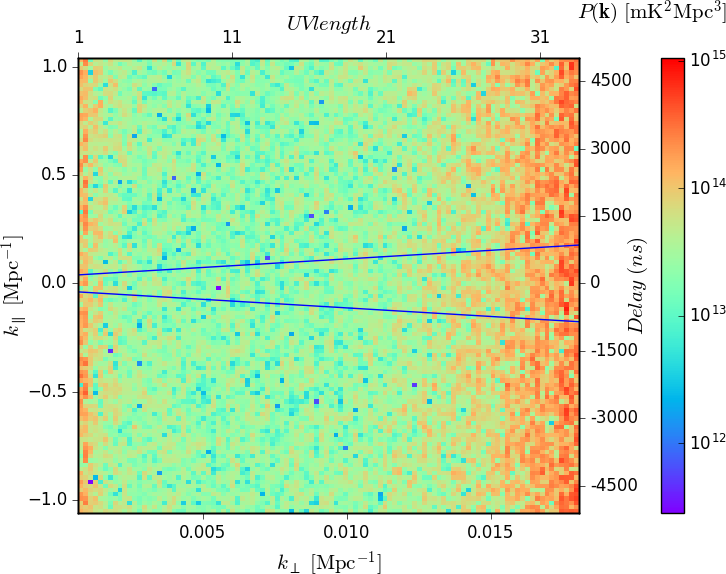}
    \end{minipage}
    \hspace{0.2in}
    \begin{minipage}[t]{\columnwidth}
        \centering
        \includegraphics[width=\columnwidth]{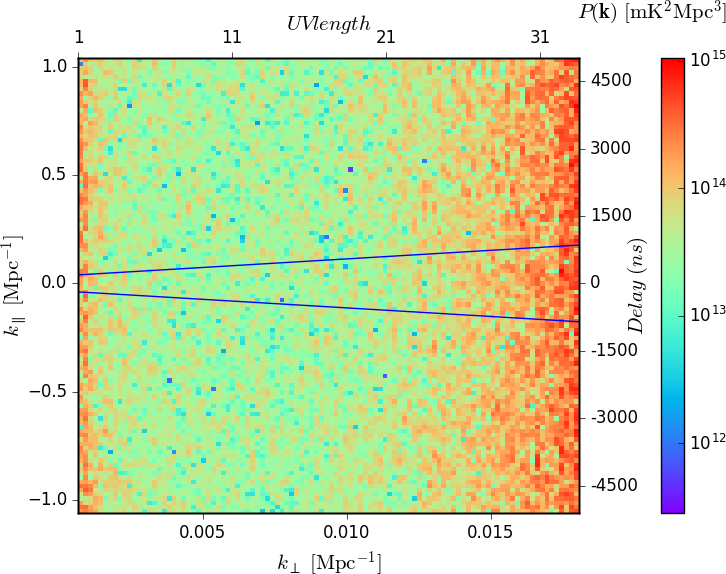}
         \vspace{0.05in}
    \end{minipage}
   
       \begin{minipage}[t]{\columnwidth}
        \centering
        \includegraphics[width=0.9\columnwidth]{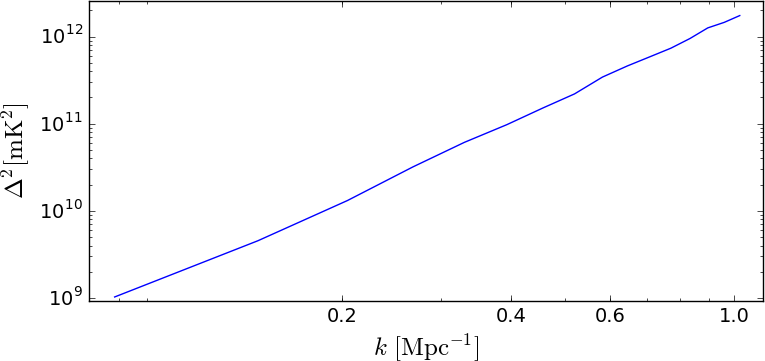}\\
    \end{minipage}
    \hspace{0.2in}
    \begin{minipage}[t]{\columnwidth}
        \centering
        \includegraphics[width=0.9\columnwidth]{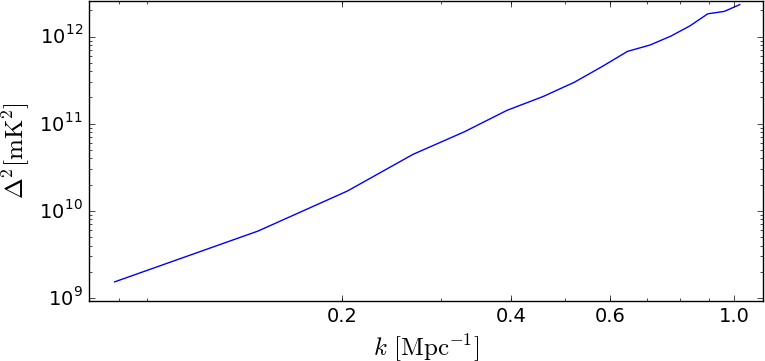}\\
     \end{minipage}

    \caption{Noise-only power spectra generated from 4\,h of simulated noise-only observations, and $\Delta^2(k)$ values obtained from them. The visibilities in the 4\,h observations  are replaced with noise values drawn from a Gaussian distribution with zero mean, and standard deviation obtained from the histograms of OVRO-LWA thermal noise   in Figure \ref{noise_values}. These simulated noise-only observations are used to generate the power spectra shown. Different  distributions are used for XX and YY polarization noise visibilities,  producing separate power spectra for XX polarization (left column) and YY polarization (right column). The top row shows the power spectra, the bottom row plots $\Delta^2(k)$ values obtained from  Cosmic Dawn window within each power spectrum (the  regions from which the values are obtained are described in Appendix \ref{appendix_regions}).}
        \label{noise_sim}
\end{figure*}

\begin{figure*}
    \begin{minipage}[t]{\columnwidth}
        \centering
        \includegraphics[width=\columnwidth]{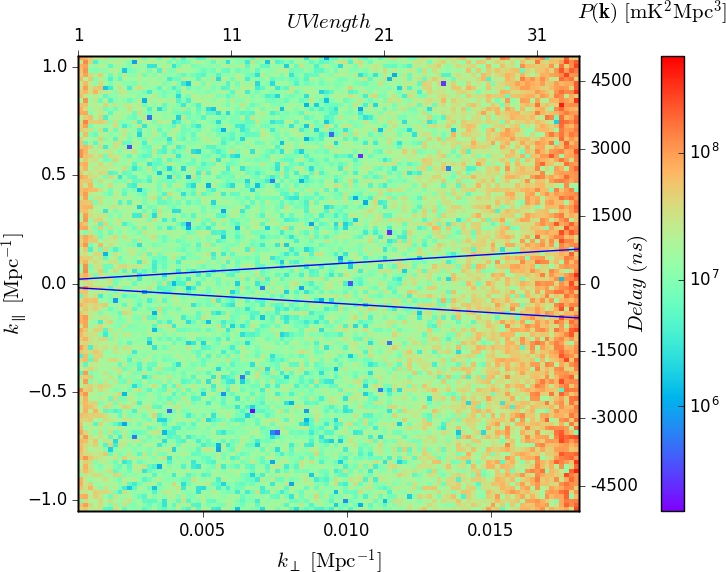}

    \end{minipage}
    \hspace{0.2in}
    \begin{minipage}[t]{\columnwidth}
        \centering
        \includegraphics[width=\columnwidth]{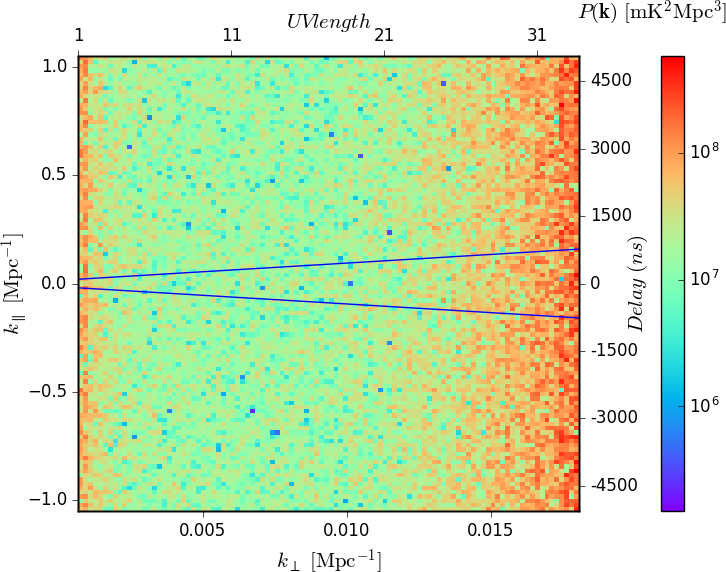}

         \vspace{0.05in}
    \end{minipage}
   
       \begin{minipage}[t]{\columnwidth}
        \centering
        \includegraphics[width=0.9\columnwidth]{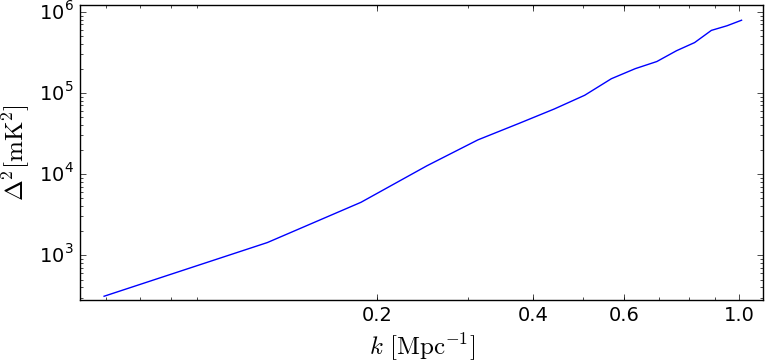}\\
    \end{minipage}
    \hspace{0.2in}
    \begin{minipage}[t]{\columnwidth}
        \centering
        \includegraphics[width=0.9\columnwidth]{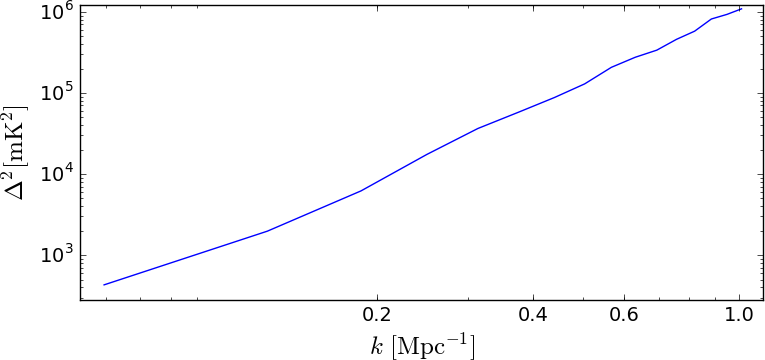}\\
         \vspace{0.05in}
    \end{minipage}

    \caption{Simulated noise-only power spectra, the same as for Figure \ref{noise_sim}, but some observations have been coherently integrated
    before power spectrum generation, as described in the text (Section \ref{coherent_integration_4hrs}).}
        \label{coherent_noise_sim}
\end{figure*}

\begin{table}
\centering
\begin{tabular}{|c||c|c|}
\hline
              &  Pol XX  & Pol YY   \\
              \hline\hline
 Real/Imag values (fit all data) &    1313.0/1312.8       &      1543.3/1543.1    \\
 \hline
 Real/Imag values (fit $\pm2000$ Jy) &    1265.8/1265.6       &      1472.2/1472.1    \\
 \hline
\end{tabular}
\caption{The standard deviations ($\sigma$) of Gaussian fits to the distributions of noise values obtained from the 4\,h of observations (Figure \ref{noise_values}).  The units are Jy. The values in the first row are obtained from fitting the histograms  using the complete x-axis range of  $\pm 8000$ Jy; the values in the bottom row are obtained from fitting in the range $\pm 2000$ Jy.}
\label{noise_stdev}
\end{table}

\begin{table}
\vspace{.1in}
\centering
\begin{tabular}{|c||c|c|}
\hline
\multicolumn{3}{|c|}{Noise power in  4\,h of  observations,}\\
\multicolumn{3}{|c|}{incoherently integrated}\\
\hline
             &  Pol XX  & Pol YY   \\
              \hline\hline
$P(\mathbf{k})$ &    $2.8 \times 10^{13}$   &  $3.8 \times 10^{13}$    \\
 \hline
$\Delta^2(k = 0.3)$  &  $4.7 \times 10^{10}$ & $6.3 \times 10^{10}$ \\
 \hline \hline
 \multicolumn{3}{|c|}{Noise power in 4\,h of observations when}\\
 \multicolumn{3}{|c|}{partial coherent integration is used}\\
 \hline
 & Pol XX & Pol YY  \\
 $P(\mathbf{k})$  &   $1.3 \times 10^{7}$   &  $1.8 \times 10^{7}$   \\

 \hline
 $\Delta^2(k = 0.3)$  & $2.3 \times 10^{4}$ & $3.2 \times 10^{4}$ \\

 \hline
\end{tabular}
\caption{Values of $P(\mathbf{k})$ and $\Delta^2(k)$ extracted from  simulated noise-only power spectra. The top two data rows are obtained from the power spectra in Figure \ref{noise_sim}, which simulates the   noise in a power spectrum generated from 4\,h of incoherently integrated observations. The bottom two data rows  are  obtained from Figure \ref{coherent_noise_sim}, which simulates the   noise in a power spectrum generated from 4\,h of partially coherently integrated observations (the integration  method is described in Section \ref{coherent_integration_4hrs}).  }
\label{psvalue_sim}
\end{table}

\begin{table*}
\begin{center}

\begin{tabular}{cccccc}
\hline
System & Total  & Redshift & $k$ for which  & Reported power \\
       &     observing time                 &          &   power reported                           &\\
\hline
OVRO-LWA \citep{2019AJ....158...84E} & 28\,h & $z = 18.4$ (73.152 MHz) & $k = 0.1$ Mpc$^{-1}$  & $\Delta^2<10^8$ mK$^2$ \\
\hline
LOFAR LBA \citep{Gehlot:2019} & 14\,h & $z = 19.8-25.2$ ($54-68$ MHz)    &  $k \approx 0.038$\,h Mpc$^{-1}$ & $\Delta^2<2.2\times10^8$ mK$^2$ \\
\hline
MWA \citep{EwallWice:2016} & 3\,h & $z = 11.6-17.9$  ($75-113$ MHz) & $k = 0.5$\,h Mpc$^{-1}$ & $\Delta^2<10^8$ mK$^2$  \\
\hline
AARTFAAC \citep{2020MNRAS.499.4158G} & 2\,h & $z = 17.9-18.6$  ($72-75$ MHz) &$k = 0.144$\,h cMpc$^{-1}$ &  $\Delta^2<7\times10^7$ mK$^2$  \\
\hline

This work & 4\,h & $z = 25-31$  ($43-54$ MHz) & $k = 0.3$\,Mpc$^{-1}$ & $\Delta^2<2\times 10^{12}$ mK$^2$  \\
\hline
\end{tabular}
\caption{Summary of 21-cm power spectrum limits obtained from observations using different telescopes. }
\label{tablimits}
\end{center}
\end{table*}

\subsection{Discussion}
\label{power_discussion}
Figure \ref{power_spectrum} demonstrates that wedge-type power spectra can be generated from OVRO-LWA observations. The power spectra show good isolation of foregrounds within the horizon, although there is some spillover. 
Our power values are  high compared to those reported by  \citet{2019AJ....158...84E}, who obtained a value of  $\Delta^2(k) \approx 10^{8}$\,mK$^2$ at $k = 0.1\,\mathrm{Mpc}^{-1}$, compared to our value of $\Delta^2(k) \approx 10^{12}$ mK$^2$ at $k = 0.3\,\mathrm{Mpc}^{-1}$.  Other telescopes also report lower power; a summary of these is shown in Table \ref{tablimits}. Note that other experiments were  using different observing frequencies,  total  observing time, and  telescopes. One reason for the higher power that we obtain is due to our lower observing frequency. At a lower frequency, foreground emission is brighter \citep{2020arXiv201103994S}, so that systematics and foreground leakage will raise the level of the power spectrum generally.


A visual inspection of the power spectra shows that horizon spillover can be separated into two distinct regions, based on the amplitude of the spillover and its extent beyond the horizon. The regions are depicted in  Figure \ref{plateau}. Both regions exist at all  $k_\perp$ values. Region 1 is bounded by the horizon and the dashed line; it  extends for approximately $\Delta k = 0.04\,\mathrm{Mpc}^{-1}, \Delta \tau = 192\, \mathrm{ns}$, beyond the horizon, and is commonly seen in other power spectra \citep[e.g.][]{2013ApJ...768L..36P, 2015ApJ...804...14T, 2015ApJ...807L..28T}. Spillover in Region 1 could be due to the window function applied to the delay transform, beam chromaticity, intrinsic spectral structure in foreground emission,  RFI, and calibration errors \citep{2020ApJ...890..122K} . The use of a window function  on the delay transform is essential  to avoid spectral leakage  due to the non-periodic nature of the baseline visibilities.  The kernel width of the Blackman Harris window we apply to our delay transforms is $288$\,ns or $\Delta k_\parallel = 0.06$, indicating it is a substantial contributor to Region 1 spillover.

Region 2 extends from Region 1 to the dashed-dot line; it extends for $\Delta  k = 0.12\,\mathrm{Mpc}^{-1}, \Delta \tau = 577\, \mathrm{ns}$ beyond Region 1. The reason for the spillover in Region 2 is not known but we suggest it is due to foreground power scattered by systematics that have not been removed by data quality checks or calibration.  
Unremoved systematics are also likely  responsible for the faint lines that can be seen in the upper part of the power spectrum. These are highlighted in Figure \ref{plateau}, upper section ("Linear features"). The  lines  begin near $k_\perp = 0, k_\parallel = 1$ and drop linearly and shallowly across the power spectrum to $k_\perp = 0.018, k_\parallel = 0.18$. 

Systematics  can  have many causes, including unmodelled diffuse emission, an inadequate beam model and/or source model for calibration, RFI, cable reflections, cross-coupling and mutual coupling, which we discuss as follows.

Models of point sources  are incomplete or approximate at  OVRO-LWA frequencies. We generated calibration models using the VLA Low-Frequency Sky Survey at 74 MHz, calculating  source flux densities at 48\,MHz by assuming a spectral index of -0.7, and assuming the location of emission at 48\,MHz is the same as it is for 74\,MHz. These assumptions likely lead to  inaccurate models.  Diffuse emission also exists within our observations, but we have not modelled it for calibration. Using more  accurate source models, and including diffuse emission  using a Global Sky Model \citep{2008MNRAS.388..247D, 2017MNRAS.469.4537D}, will improve calibration.

 OVRO-LWA observations are affected by cross-coupling of antenna signals,   where cables are in close proximity as they enter backend hardware processing units. It is known to impact 509 baselines, and these are ignored when generating power spectra, but they may not be the only ones affected. A more detailed investigation is needed.  

Electromagnetic  interaction between neighboring antennas (mutual coupling) may be present in OVRO-LWA observations, including coupling between the X and Y dipoles on the same OVRO-LWA stand, leading to polarization leakage. A study of mutual coupling in the LWA reported that it exists, but has no consistent positive or negative effect \citep{2011ITAP...59.1855E}, however more investigation of these effects is needed for OVRO-LWA.

Electrical signals travelling via cable from an antenna to backend  components may reflect from those components,  introducing sinusoidal structure  in the antenna frequency response.  For the most part, cross-correlations between antennas (i.e. visibilities) do not suffer this systematic as the noise waves along a pair of cables are not correlated. However,  systematics due to cable reflections have been reported in other power spectra \cite[e.g][]{2016ApJ...833..102B} and may be present in ours. The linear features in Figure \ref{plateau} may   be due to cable reflections, but in that case we expect the features to be horizontal. 

RFI flaggers such as  AOFlagger \citep{2012A&A...539A..95O} will provide better RFI detection compared to the simple algorithm implemented within RTS. Flaggers will be implemented in the future within the OVRO-LWA pipeline. More sophisticated beam models can be generated through the use of  simulation packages (e.g. FEKO\footnote{https://altairhyperworks.com/product/Feko/Applications-Antenna-Design}, NEC\footnote{https://www.qsl.net/4nec2/}, CST\footnote{https://uk.mathworks.com/products/connections/product\_detail/cst-microwave-studio.html}), direct measurements of the radiation field of the antenna (for example by using drones \citep{6929024} or satellites \citep{2015RaSc...50..614N}), or by observing radio sources transiting through the fringes of the interferometer  \citep{2016ApJ...820...51P}.

Methods for the detection, analysis, and removal of systematics in power spectra are being developed by other groups and  reported in the literature. For example, \citet{2020ApJ...890..122K} demonstrates the usefulness of analysing antenna-based gains in delay and fringe-rate space,  using gain-smoothing  to suppress  delay modes that appear to be abnormal, and using temporal analysis and temporal smoothing of gains using autocorrelations. Cross-coupling can be removed by applying a high-pass filter in fringe rate space \citep{2020ApJ...888...70K}.
 Window functions can be analysed for their contribution to   horizon spillover, and  alternative window functions selected \citep{2020MNRAS.tmp.1168L}. Use of these and other methods will improve the quality of power spectra generated from OVRO-LWA.

\begin{figure*}
\centering
\includegraphics[scale=0.8]{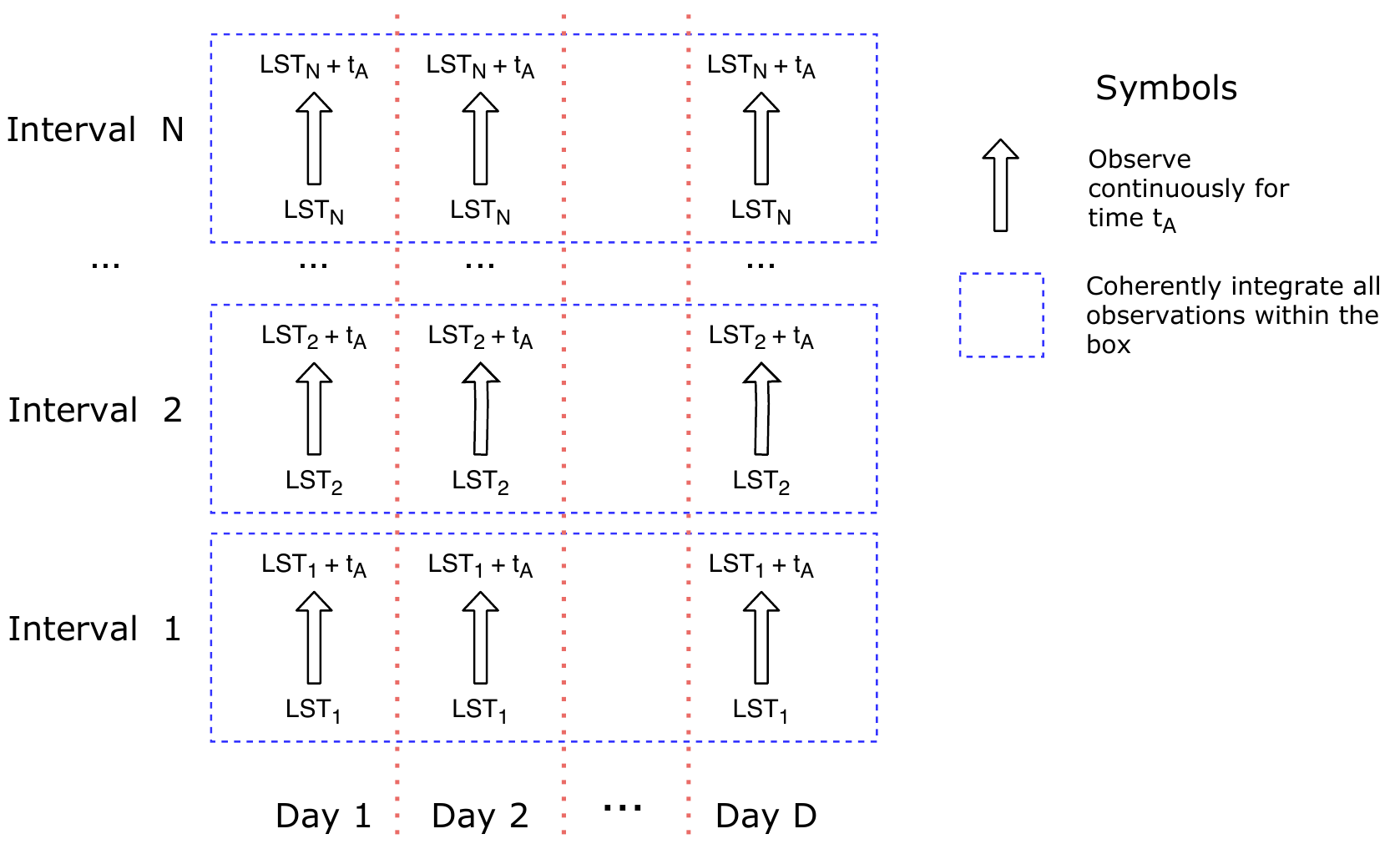}
\caption{A scheme for making observations with a drift-scan interferometer that can  be partially coherently integrated. Observe for $N$  short time Intervals  ($LST_\mathrm{i} \rightarrow LST_\mathrm{i+t_A}$) each day, repeat the observations over several days, and coherently integrate all observations within each Interval across all days. The time range  $t_\mathrm{A}$ is the time-averaging limit. The resulting $N$ coherently integrated observations must be combined incoherently.}
\label{coh_int}
\vspace{0.2in}
\end{figure*}

\section{The sensitivity of OVRO-LWA}
\label{sensitivity_analysis}

We now turn to the analysis of the sensitivity of power spectra generated from OVRO-LWA using the delay spectrum method, and investigate the sensitivity that can be achieved using a large number of  observations and a different  integration strategy.

We determine the level of  noise in OVRO-LWA observations -- ``noise'' includes thermal instrumental and sky noise.  We use simulations to generate power spectra containing only noise; these allow us to find the noise floor in our power spectra from observations (Figure \ref{power_spectrum}). We then  investigate the noise floor in power spectra generated from observations that are partly coherently integrated before power spectrum generation.  Coherent integration should significantly reduce the noise level in power spectra.

\subsection{Measuring the OVRO-LWA  thermal noise}
\label{measure_noise}
The noise in our observations is extracted from observed  visibilities using a simple data-differencing technique (\cite{2010A&A...522A..67B, 2019MNRAS.483.2269C}) that subtracts visibilities in adjacent frequency channels. We assume that subtracting visibilities at slightly different frequencies will subtract out the sky signal and systematics, leaving only subtracted noise values.   From the distribution of these noise values we obtain the statistical properties of the noise, from which we can generate  noise-only power spectra using simulations. The noise level in these power spectra is interpreted as the level of noise that would be present in power spectra generated from telescope observations.
For visibility differencing, we use all the visibilities in all the calibrated observations used for the power spectra in Figure \ref{power_spectrum}, ignoring flagged visibilities.

Figure \ref{noise_values} show histograms of the  values obtained from frequency differencing, for XX and YY polarizations, and for the real and imaginary parts of the  values. The distributions are  close to Gaussian, with a zero mean,  giving us confidence that we have in fact extracted mostly thermal noise. However, there are  wings in the data distribution  indicating an excess of values around $\pm 3000$\,Jy, and the data peak is higher than the fitted peak. To investigate these, we generated images of the visibility differences from one of the OVRO-LWA observations. The images mostly contained noise, but  showed that Virgo A, and the RFI sources on the horizon (Figure \ref{the_sky}), were not always completely removed, leading to an excess of values at the location of the wings.

The standard deviations $\sigma$ of the Gaussian fits to the noise histograms are listed in Table \ref{noise_stdev}, top row. These fits include the wings previously mentioned; if the wings were not present, we would expect a narrower distribution and smaller $\sigma$ values.  To estimate what these would be, we fit a Gaussian to the noise histograms in the x-axis range $\pm 2000$\,Jy only. The standard deviations for these fits are in Table \ref{noise_stdev}, second row, and produce $\sigma$ that are about 5\% smaller than those resulting from fitting the entire x-range.

To be conservative, we  use the higher values of $\sigma$   for our noise simulations.

\subsection{Measuring the noise floor in the power spectra generated from 4\,h of telescope observations (incoherent integration)}
\label{noise_only_sim}

We simulate  noise-only observations by replacing all the visibilities in all the 4\,h of telescope observations with noise values drawn from a Gaussian distribution with zero mean and standard deviation listed in Table \ref{noise_stdev}. Baselines that were flagged in the  observations are flagged in the  simulated noise-only observations, but flagged channels are filled in with a noise value. A ``noise-only'' power spectrum is generated from the 4\,h of simulated noise-only observations; these power spectra are shown in Figure \ref{noise_sim}, and the power values extracted from them are listed in Table \ref{psvalue_sim}. The power spectra generated from  observations have a value of $P(\mathbf{k}) \approx 5\times 10^{14}$ in the Cosmic Dawn window (Table \ref{psvalue}), but the noise-only power spectra have  $P(\mathbf{k}) \approx 3\times 10^{13}$ in the same region,   demonstrating that there are  systematics in the observed data. 

All the power spectra presented so far were generated using incoherent integration; we now turn to experiments with coherent integration.

\subsection{Measuring the noise floor in the power spectra generated from 4\,h of telescope observations (if coherent integration is used)}
\label{coherent_integration_4hrs}


We follow a method suggested by \cite{2014ApJ...793...28P}, where observations are made within several short time intervals $t_A$, repeated over many days. Each time interval begins at the same LST each day, and all observations made  during time $t_A$,   at the same LST, may be coherently integrated.  The scheme is depicted in Figure \ref{coh_int}. $t_A$ is the ``time-averaging limit'';  the length of time over which observations may be safely coherently integrated (see Appendix \ref{integration_time_limit}). By following this method, $N$ coherently integrated observations are produced, which must then be incoherently integrated. We refer to this scheme as ``partial coherent integration''.

We note that it may be possible to apply this mix of incoherent/coherent integration to the 4\,h of telescope observations that were used to generate the power spectra  in  Figure \ref{power_spectrum}, but it depends on the LSTs of the observations, as they must fit into appropriate time intervals as described above.  An investigation of this possibility, and publication of  power spectra that may result, is   reserved for future work.

To demonstrate the use of the partial coherent integration method, and the noise reduction that can be achieved, we use the 4\,h of simulated noise-only observations from the previous Section, altering the observation LSTs so we can apply the  method. At 48.342\,MHz, $t_\mathrm{A} = 36$\,s (Appendix \ref{integration_time_limit}), so we simulate observing at 4 different LSTs ($N = 4$, Figure \ref{coh_int}) for 36\,s each, over 100 days ($D =100$).
It is assumed that the thermal noise in the telescope does not vary over the 100 days, so that all noise values can be obtained from the distributions in Figure \ref{noise_values}.  
 
 Figure \ref{coherent_noise_sim} shows the  power spectrum generated as a result, and Table \ref{psvalue_sim} (bottom two rows) shows the average $\Delta^2(k)$  extracted from the power spectra. These show that we can reduce the noise level in the power spectrum by a factor of $\approx 10^6$, when using partial coherent integration, instead of fully incoherent integration.

\subsection{Measuring the noise floor in a power spectra generated from 3000\,h of telescope observations (partial coherent integration)}

Following on from the previous experiment, we  obtain the noise level in power spectra generated from  3000\,h   of partially coherently integrated observations. These will be compared with the power in 21-cm fluctuations in the next Section.

We set  the number of observing days to $D = 750$, or 2 years. An observing schedule over 2 yrs may be  difficult to sustain, but feasible. For OVRO-LWA it will also generate 169\,TiB of data per sub-band, or 3.5 \,PiB in total, also feasible using  a large computer cluster.  We fix the observing time per day to 4\,h,  so as to avoid transits of the Galactic center and other bright sources. We assume there are no transits of local bodies. We again assume that the noise properties of the telescope do not change over the 750 days, and that there are also no  changes in telescope performance, availability, and reliability. 
The time-averaging limit $t_\mathrm{A}$ varies by frequency, so the number of Intervals $N$  (Figure \ref{coh_int}) must be different for each sub-band, to accumulate 4\,h per day.  At  $48.342$\,MHz with $t_\mathrm{A} = 36$\,s, $N$ = 400;  at 30 MHz with  $t_\mathrm{A} = 54$\,s,  $N$ = 267. 

We use a frequency range of 37.848-84.912\,MHz ($z = 15.7-36.5$), dividing it  into 18 sub-bands, each of width of 5.232\,MHz. Each sub-band overlaps the next by 2.616\,MHz, and are separated in frequency by 2.616\,MHz. The center frequency is used to determine the redshift of the sub-band.

Observations at different frequencies will have different noise properties. We assume that the noise distribution will be Gaussian as depicted in Figure \ref{noise_values}, but that  the  $\sigma$ of the noise distribution will vary by frequency.  Since  the noise  distributions were derived from observed visibilities, we assume that $\sigma$ varies as a power law with  exponent -0.7, as was assumed for the source flux densities in Section \ref{data_description}. Therefore, $\sigma_\nu$ for the noise distribution at frequency $\nu$ is  calculated as:
\begin{equation}
    \sigma_\nu = \sigma_{48.324}*(\nu/48.324\times 10^6)^{-0.7}.
\end{equation}

 A power spectrum is generated  for each sub-band, and  the value of $\Delta^2(k)$ at $k = 0.3$ extracted from the power spectra. This gives a $\Delta^2(k)$ noise level for a range of redshifts, which can be compared against expected power based on theoretical models of the cosmological 21-cm signal, in the next Section.

Generating  3000 hours of noise-only observations is very compute intensive, and generates 3.5\,PiB of data, so we implement it in a GPGPU. Since the observations all contain the same number of baselines and channels (none are flagged), a GPGPU thread is created for every channel in every baseline, and each thread generates and manipulates noise values  ``on the fly'', avoiding the need to store the values.  A full description of the GPU implementation will be published elsewhere.



\section{Investigating Cosmic Dawn using OVRO-LWA Observations}
\label{cosmic_dawn_with_ovro}

From the simulations described in the previous Section, we obtain the noise power in OVRO-LWA 21-cm power spectra, generated from 3000\,h of partially coherently integrated observations, over a range of redshifts.  If noise is the only contributor to power in the Cosmic Dawn window, apart from the 21-cm fluctuations, then the noise power determines whether a 21-cm detection is possible. In this Section we assume that this is so, and discuss in detail what may be detected according to theoretical models. However,  ensuring that noise is the only other source of power in the Cosmic Dawn window requires the elimination of all systematics that we have discussed above. This may difficult at present, as  methods that exist for removal of systematics are still being proven \citep[e.g. see][]{2020ApJ...890..122K}, but it is plausible in the future, at least, we suggest, to a level where systematic effects are insignificant. We base the following discussion  on this ``best-case'' scenario.

Being a high-redshift instrument, OVRO-LWA is by design observing the 21-cm signal from the epoch of  primordial star and black hole formation. Thus, observations with this instrument  could be used to infer properties  of the first sources as well as to constrain exotic physics at Cosmic Dawn. 

\begin{figure}
\includegraphics[width=\columnwidth]{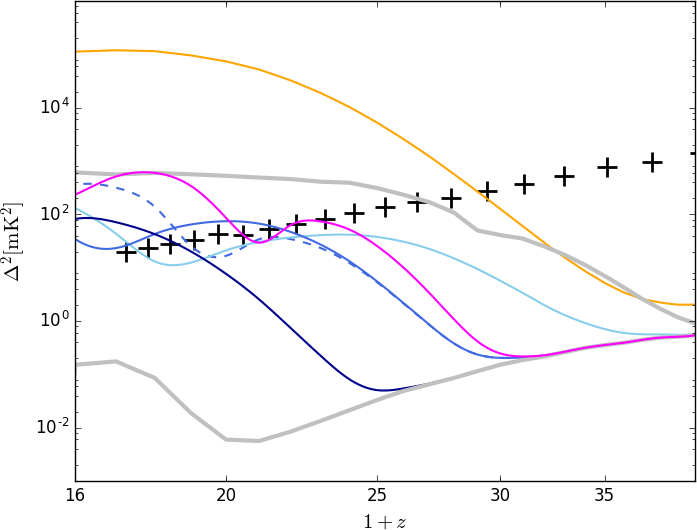}

\caption{The simulated noise level of OVRO-LWA from 3000\,h of partially coherently integrated observations at $k = 0.3$\,Mpc${^{-1}}$  (crosses), plotted on top of a selection of 21-cm models (note the x-axis is in log scale). We show standard models with  $V_\mathrm{c} = 4.2$\,km/s (light blue), $24.2$\,km/s (blue) and $52.1$\,km/s (dark blue) for fixed values of $f_\ast=3\%$, $f_\mathrm{X} = 0.07$, $\alpha = 1.5$, $\nu_\mathrm{min} =0.1$\,keV (the EoR parameters, although they do not matter much at these redshifts, are chosen to be $R_\mathrm{mfp} = 30$\,Mpc and $\tau = 0.064$). For comparison we also show the case of higher $f_\mathrm{X} = 1.7$ for $V_\mathrm{c}  = 24.2$\,km/s (dashed blue). We also show the model with maximum signal to noise in the LEDA band (magenta, out of examined 11164 modes). This case has the following values of the astrophysical parameters: $f_\ast=50\%$, $V_\mathrm{c} = 35.5$\,km/s,  $f_\mathrm{X} = 1$ with $\alpha = 1.5$ and $\nu_\mathrm{min} =0.1$\,keV, $R_\mathrm{mfp} = 40$\,Mpc and $\tau = 0.07$. We also show the envelope of all possible signals in the standard astrophysical scenario \citep[thick gray lines,][]{2018MNRAS.478.2193C}. Finally, we show an {\it external radio background}  model (orange) that has the maximum signal to noise for LEDA: $f_\ast=12\%$, $V_\mathrm{c} = 4.2$\,km/s,  $f_\mathrm{X} = 0.01$ with hard X-ray SED, $\tau = 0.056$ and the excess radio background of 1.2\% over the CMB at 1.42\,GHz  \citep[from][]{2019arXiv190202438F}.
}
\label{LEDA_PS}
\end{figure}

The noise power $\Delta^2(k=0.3)$ from 3000\,h of partially coherently integrated observations   is plotted as crosses in Figure \ref{LEDA_PS}, for the redshift range  $17 \le 1+z \le 38$. To demonstrate that  OVRO-LWA is able to detect the 21-cm signal at these redshifts,  the Figure also contains $\Delta^2(k)$ for  a sample  of theoretical models, including the envelope of all possible signals in the standard astrophysical scenario  \citep[as explained below, and in][]{2017MNRAS.472.1915C}, and an exotic physics case which was proposed to explain the deep EDGES absorption feature \citep[here for illustrative purposes we show a model with enhanced radio background over the CMB, see][]{2019arXiv190202438F}. The models are described in the Figure caption. Some of the standard astrophysical models, along with the exotic scenario, are well above the sensitivity of OVRO-LWA at these redshifts, indicating that   OVRO-LWA is expected to be sensitive enough to dig deep into the discovery space of the predicted 21-cm signals.  

Let us first examine the standard astrophysical case, i.e., models that assume the CMB as  background radiation, with a conventional cold dark matter scenario,  and hierarchical structure formation. The astrophysical model contains seven parameters  \citep[see][for a recent summary of the astrophysical modeling]{2019arXiv191006274C}, including: star formation  efficiency $f_\ast$, minimum halo mass suitable for star formation (or, equivalently, minimum circular velocity of such halos, $V_\mathrm{c}$); X-ray efficiency of sources compared to their present-day counterparts $f_\mathrm{X}$; the spectral properties of X-ray sources (namely, the slope of X-ray spectral energy distribution,  $\alpha$, and the minimum frequency,  $\nu_\mathrm{min}$). We  model the process of reionization using two more free parameters (the total CMB optical depth, $\tau$,  and the mean free path of UV photons, $R_{\rm mfp}$); however, reionization does not play a significant role at  the high-redshift regime probed by OVRO-LWA. 

Astrophysics is in its simplest form at  high redshifts accessible using OVRO-LWA.  With first stars forming in small and rare dark matter halos, and owing to the absence of massive galaxies and AGN, there are relatively few processes that affect the early 21-cm signal. Thus, constraining the 21-cm signal at high redshifts with OVRO-LWA  would offer  one of the purest probes of primordial star formation. The most prominent feature of the 21-cm signal from this epoch is the high-redshift peak in the power spectrum imprinted by the non-homogeneous  Ly-$\alpha$  field \citep{2005ApJ...626....1B}.   The amplitude and central frequency of the peak  depend on just two parameters in our case ($f_\ast$ and $V_\mathrm{c}$),   and the peak is above the sensitivity of OVRO-LWA for models with efficient star formation in small dark matter halos.  Thus, using 3000\,h of observations, OVRO-LWA can probe properties of the first star forming halos, star formation efficiency, and put constraints on stellar feedback. The second effect that can be constrained by OVRO-LWA  is the onset of the IGM heating by the first X-ray sources. Although in some scenarios X-ray heating is delayed, in many of the examined  models X-ray sources turn on early and could imprint heating fluctuations in the gas temperature in the OVRO-LWA band. Examining a set of $\approx 11000$ models with different combinations of astrophysical parameters, we find that the strongest signal to noise in the OVRO-LWA band has a model with efficient star formation ($f_\ast = 50\%$), with a minimum mass of star forming halos of $M_\mathrm{h} \approx 3\times 10^8\,M_\odot$,  and close to present-day X-ray heating efficiency 
(magenta curve in Figure \ref{LEDA_PS}). For this model  both the Ly-$\alpha$ peak at $z\approx 22$ and the X-ray peak at $z\approx 16.5$ can be measured by OVRO-LWA.

For the other standard models shown in Figure \ref{LEDA_PS} (shades of blue), either the X-ray or the Ly-$\alpha$ peaks can be detected. We compare the case of dark matter halos with $V_\mathrm{c} = 4.2$, 24.2 and 52.1\,km/s, for  otherwise fixed parameters. The Ly-$\alpha$ peak for $V_\mathrm{c} = 4.2$\,km/s occurs at high redshifts where the sensitivity of OVRO-LWA is weak, and, thus, cannot be observed. However, the rise of the power imprinted by the temperature fluctuations can be measured in the highest frequency channels. For the other two cases ($V_\mathrm{c} =24.2$  and 52.1\,km/s) the Ly-$\alpha$ peak is at lower redshifts and happens to be above the noise curve, and, thus, is detectable. However, for these models the X-ray peak is shifted to higher frequencies out of the OVRO-LWA band. Finally, varying X-ray parameters can also play a role; compare the case of $V_\mathrm{c} = 24.2$\,km/s with weak X-ray sources (solid blue, $f_\mathrm{X} = 0.07$) to a case with the same $V_\mathrm{c}$ but stronger X-ray heating ($f_\mathrm{X} = 1.7$, dashed blue). Because the effects of heating and Ly-$\alpha$ coupling on the 21-cm signal anti-correlate, more efficient X-ray heating results in a lower Ly-$\alpha$ peak which is difficult to detect, but also in an earlier and stronger X-ray peak that shifts into the OVRO-LWA band. 


If the anomalously deep global signal detected by EDGES Low-Band is astrophysical, the 21-cm power spectrum is expected to be boosted by a few orders of magnitude   \citep[e.g.,][]{Fialkov:2018, 2019arXiv190202438F,2020MNRAS.499.5993R}. 
The exact amount of the enhancement relative to a  standard scenario with the same astrophysical parameters, depends on the underlying theory which explains the EDGES observation. Owing to this boost, it would be much easier to detect such  signals. As an illustration, we show one example of such a signal in  Figure \ref{LEDA_PS} (orange line) for a theory in which the 21-cm signal is enhanced due to an excess radio background  over the CMB at high redshifts \citep[from][]{2019arXiv190202438F}.

\begin{figure}
\includegraphics[width=\columnwidth]{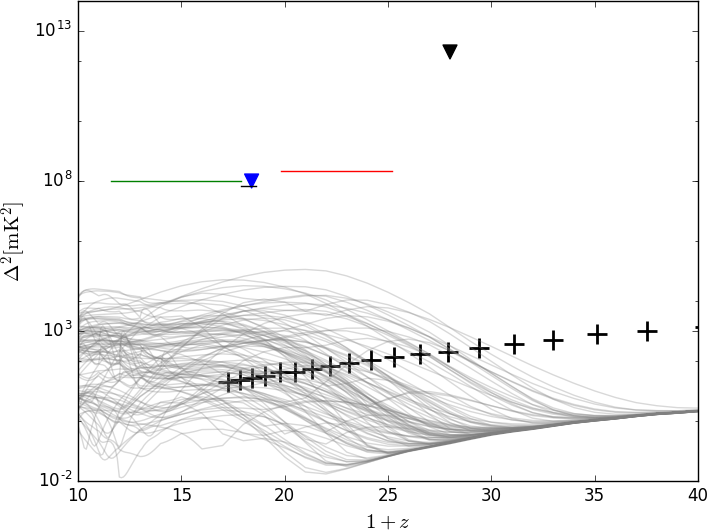}
\caption{We show our limit of $2\times 10^{12}$\,mK$^2$ at $z = 28$ and at $k = 0.3$  Mpc$^{-1}$ obtained from 4\,h of observations (black triangle) in the context of limits from other high redshift 21-cm probes including LOFAR (NCP, red line) derived from 14\,h of data at 54-68 MHz ($z = 19.8-25.2$)  at $k\approx 0.038$\,h Mpc$^{-1}$ \citep{Gehlot:2019}, MWA (green line) derived from 3\,h of data at 75-113\,MHz (corresponding to $z = 11.6-17.9$) at $ k = 0.5$\,h\,Mpc$^{-1}$ \citep{EwallWice:2016},  OVRO-LWA (blue triangle)  obtained using 28\,h of data at 73.152\,MHz ($z = 18.4$) at $ k = 0.1$\,h Mpc$^{-1}$ \citep{2019AJ....158...84E}, and AARTFAAC (black line) obtained using 2\,h of data at $72-75$ MHz ($z = 17.9-18.6$) at $ k = 0.144$\,h\,cMpc$^{-1}$ \citep{2020MNRAS.499.4158G}. We also show the predicted OVRO-LWA sensitivity curve from the 3000\,h of  noise data (black crosses) obtained as described in the previous Section, and a hundred randomly selected 21-cm power spectra modeled with extra radio background, from \citet{2019arXiv190202438F}. 
}
\label{compare_3000}
\end{figure}

Finally, in Figure \ref{compare_3000}, we  compare the sensitivity level obtained from 3000\,h of observations, with the power  obtained from the 4\,h of observations (Section \ref{results}), and the power obtained from other telescopes as reported in the literature (Table \ref{tablimits}). We include a sample of models with excess radio background \citep{2019arXiv190202438F}. 


\section{Conclusions}
\label{summary}
We have demonstrated that the delay spectrum method can be used to generate power spectra from OVRO-LWA observations, although they contain systematics that prevent us from reaching the  level of 21-cm fluctuations  in the Cosmic Dawn window. Many systematics can influence power spectrum sensitivity,  including: inaccurate sky models for calibration (including unmodelled diffuse mission), undetected RFI, inaccurate beam model (shape and chromaticity), gain calibration errors, cross-coupling and mutual coupling, and cable reflections; all of which are present, or likely to be present, in our power spectra.  Mitigating systematics in the context of 21-cm power spectra is an active area of research, and we will borrow promising techniques developed for use with other telescopes, such as the MWA, HERA, SKA, and LOFAR.

We have also demonstrated that a compact, imaging interferometer is able to generate a 21-cm power spectrum, indicating that such interferometers have a role to play in detecting the 21-cm signal in power spectra.

We find that the power level in 21-cm fluctuations  at 48.324\,MHz is $\Delta^2(k) \approx 2 \times 10^{12}\,\mathrm{mK}^2$ at $k = 0.3\, \mathrm{Mpc}^{-1}$, after incoherently integrating  4\,h of observations. Simulations of OVRO-LWA thermal noise indicate that the  noise floor in the power spectra is  $\Delta^2(k) \approx 5 \times 10^{10}\  \mathrm{mK}^{2}$. This level is too high to detect  21-cm fluctuations at Cosmic Dawn,  but we may take the value of $\Delta^2(k) \approx 2 \times 10^{12}\,\mathrm{mK}^2$ as an upper limit. Simulations show that if we were able to implement our partial coherent integration method, and apply it to 3000\,h of observations, the noise floor would drop to $10^2$\,mK$^2$, allowing for a detection according to theoretical models.

However the noise simulations are simplistic. Firstly, properties of the noise were obtained from visibility differencing.  A more rigorous treatment would involve precise measurements and simulations of the noise in telescope components \citep[e.g. as for the MWA,][]{2020arXiv200305116U} followed by an analysis of how this converts  to  noise power $\Delta^2(k)$ in a 21-cm power spectrum. Secondly, observing over a period of many years,  to obtain observations that can be coherently integrated, will require that the telescope remains stable, along with atmospheric and environmental conditions, and that observations are not affected by transient events and unwanted objects passing through the field of view. It is unlikely these can all be controlled, so some observations may need to be discarded, increasing the total observing time to compensate, or their effects will have to be mitigated in the data. These issues will have to be planned for.

We also mentioned that the delay spectrum technique is not suitable for the frequency range and baseline lengths that we have used. Long baselines can be excluded from power spectra, which will result in a shortened $k_\perp$ range. Restricting the frequency range is problematic as it will result in  few frequency channels being used for power spectrum generation. For example, using Equation \ref{appr_limit} and again interpreting $\ll$ as ``one-half'', and limiting baseline lengths to 50 meters, the allowable bandwidth is 1.6 MHz. This means only 70 OVRO-LWA frequency channels (of width 24 kHz) can be used, compared to the 436 that were used for the power spectra in this paper, resulting in  fewer  $k_\parallel$ modes available for analysis.

The  redshift range over which a power spectrum can made at high redshift, so that it measures a static Universe,  is uncertain, and will not be determined until we have more knowledge of the actual rate of change of  state of the early Universe, perhaps from global 21-cm experiments.  The range will depend on the redshift of the observation; at $z \approx 9$, \citet{2013ApJ...768L..36P} suggest a redshift range of $\Delta z \approx 0.5$ as being acceptable. At $z \approx 28$, as used here, a redshift range of 0.5 corresponds to a frequency range of 0.8 MHz, which allows for only 33 OVRO-LWA channels to be used for power spectrum generation. This and the discussion in the previous paragraph indicate that the OVRO-LWA frequency resolution should be increased for future work.

We may also expand future work to include other methods of power spectrum generation, such as removing foregrounds from observations using foreground models \citep[e.g. as in][]{Gehlot:2019},  or adopting the m-mode analysis used by \citet{2018AJ....156...32E}. However, many of the issues mentioned above still remain, and should be addressed before doing so.

\section*{Acknowledgements}

The authors thank the referee, whose constructive and helpful comments enabled us to improve the quality of the paper.
 AF was supported by a Royal Society University Research Fellowship. HG was supported in part through a visiting scientist position at the Smithsonian Astrophysical Observatory (SAO).
Phases of the LEDA project have been supported by NSF grants AST/1616709, AST/1106059, OIA/1125087, and PHY/0835713.  The Owens Valley Radio Observatory (OVRO) is operated by the California Institute of Technology. The OVRO-LWA project was enabled by the kind early donation of Deborah Castleman and Harold Rosen. LG thanks members of the SAO Receiver Lab for their generous collaboration and OVRO site staff for advice kindly offered, critical and routine assistance, and collegial spirit.   

\section*{Data Availability}

The data used to generate the results reported in this paper are not publicly available, but  will be shared, on reasonable request to the corresponding author.



\bibliographystyle{mnras}
\bibliography{main} 

\begin{thebibliography}{}
\makeatletter
\relax
\def\mn@urlcharsother{\let\do\@makeother \do\$\do\&\do\#\do\^\do\_\do\%\do\~}
\def\mn@doi{\begingroup\mn@urlcharsother \@ifnextchar [ {\mn@doi@}
  {\mn@doi@[]}}
\def\mn@doi@[#1]#2{\def\@tempa{#1}\ifx\@tempa\@empty \href
  {http://dx.doi.org/#2} {doi:#2}\else \href {http://dx.doi.org/#2} {#1}\fi
  \endgroup}
\def\mn@eprint#1#2{\mn@eprint@#1:#2::\@nil}
\def\mn@eprint@arXiv#1{\href {http://arxiv.org/abs/#1} {{\tt arXiv:#1}}}
\def\mn@eprint@dblp#1{\href {http://dblp.uni-trier.de/rec/bibtex/#1.xml}
  {dblp:#1}}
\def\mn@eprint@#1:#2:#3:#4\@nil{\def\@tempa {#1}\def\@tempb {#2}\def\@tempc
  {#3}\ifx \@tempc \@empty \let \@tempc \@tempb \let \@tempb \@tempa \fi \ifx
  \@tempb \@empty \def\@tempb {arXiv}\fi \@ifundefined
  {mn@eprint@\@tempb}{\@tempb:\@tempc}{\expandafter \expandafter \csname
  mn@eprint@\@tempb\endcsname \expandafter{\@tempc}}}

\bibitem[\protect\citeauthoryear{{Anderson} \& {Hallinan}}{{Anderson} \&
  {Hallinan}}{2017}]{2017reph.conf40102A}
{Anderson} M.~M.,  {Hallinan} G.,  2017, in Radio Exploration of Planetary
  Habitability (AASTCS5). p. 401.02

\bibitem[\protect\citeauthoryear{{Anderson} et~al.,}{{Anderson}
  et~al.}{2018}]{2018ApJ...864...22A}
{Anderson} M.~M.,  et~al., 2018, \mn@doi [\apj] {10.3847/1538-4357/aad2d7},
  \href {https://ui.adsabs.harvard.edu/abs/2018ApJ...864...22A} {864, 22}

\bibitem[\protect\citeauthoryear{{Anderson} et~al.,}{{Anderson}
  et~al.}{2019a}]{2019ApJ...886..123A}
{Anderson} M.~M.,  et~al., 2019a, \mn@doi [\apj] {10.3847/1538-4357/ab4f87},
  \href {https://ui.adsabs.harvard.edu/abs/2019ApJ...886..123A} {886, 123}

\bibitem[\protect\citeauthoryear{{Anderson}, {Callister}  \&
  {Hallinan}}{{Anderson} et~al.}{2019b}]{2019GCN.24196....1A}
{Anderson} M.~M.,  {Callister} T.~A.,   {Hallinan} G.,  2019b, GRB Coordinates
  Network, \href {https://ui.adsabs.harvard.edu/abs/2019GCN.24196....1A}
  {24196, 1}

\bibitem[\protect\citeauthoryear{{Barkana}}{{Barkana}}{2016}]{Barkana:2016}
{Barkana} R.,  2016, \mn@doi [\physrep] {10.1016/j.physrep.2016.06.006}, \href
  {https://ui.adsabs.harvard.edu/abs/2016PhR...645....1B} {645, 1}

\bibitem[\protect\citeauthoryear{{Barkana}}{{Barkana}}{2018}]{Barkana:2018}
{Barkana} R.,  2018, \mn@doi [\nat] {10.1038/nature25791}, \href
  {https://ui.adsabs.harvard.edu/abs/2018Natur.555...71B} {555, 71}

\bibitem[\protect\citeauthoryear{{Barkana} \& {Loeb}}{{Barkana} \&
  {Loeb}}{2005}]{2005ApJ...626....1B}
{Barkana} R.,  {Loeb} A.,  2005, \mn@doi [\apj] {10.1086/429954}, \href
  {https://ui.adsabs.harvard.edu/abs/2005ApJ...626....1B} {626, 1}

\bibitem[\protect\citeauthoryear{{Barry} et~al.,}{{Barry}
  et~al.}{2019}]{Barry:2019}
{Barry} N.,  et~al., 2019, \mn@doi [\apj] {10.3847/1538-4357/ab40a8}, \href
  {https://ui.adsabs.harvard.edu/abs/2019ApJ...884....1B} {884, 1}

\bibitem[\protect\citeauthoryear{{Beardsley} et~al.,}{{Beardsley}
  et~al.}{2016}]{2016ApJ...833..102B}
{Beardsley} A.~P.,  et~al., 2016, \mn@doi [\apj] {10.3847/1538-4357/833/1/102},
  \href {https://ui.adsabs.harvard.edu/abs/2016ApJ...833..102B} {833, 102}

\bibitem[\protect\citeauthoryear{{Bernardi} et~al.,}{{Bernardi}
  et~al.}{2010}]{2010A&A...522A..67B}
{Bernardi} G.,  et~al., 2010, \mn@doi [\aap] {10.1051/0004-6361/200913420},
  \href {https://ui.adsabs.harvard.edu/abs/2010A&A...522A..67B} {522, A67}

\bibitem[\protect\citeauthoryear{{Bernardi} et~al.,}{{Bernardi}
  et~al.}{2016}]{2016MNRAS.461.2847B}
{Bernardi} G.,  et~al., 2016, \mn@doi [\mnras] {10.1093/mnras/stw1499}, \href
  {http://adsabs.harvard.edu/abs/2016MNRAS.461.2847B} {461, 2847}

\bibitem[\protect\citeauthoryear{{Bowman}, {Rogers}, {Monsalve}, {Mozdzen}  \&
  {Mahesh}}{{Bowman} et~al.}{2018}]{2018Natur.555...67B}
{Bowman} J.~D.,  {Rogers} A.~E.~E.,  {Monsalve} R.~A.,  {Mozdzen} T.~J.,
  {Mahesh} N.,  2018, \mn@doi [\nat] {10.1038/nature25792}, \href
  {http://adsabs.harvard.edu/abs/2018Natur.555...67B} {555, 67}

\bibitem[\protect\citeauthoryear{{Bradley}, {Tauscher}, {Rapetti}  \&
  {Burns}}{{Bradley} et~al.}{2019}]{Bradley:2019}
{Bradley} R.~F.,  {Tauscher} K.,  {Rapetti} D.,   {Burns} J.~O.,  2019, \mn@doi
  [\apj] {10.3847/1538-4357/ab0d8b}, \href
  {https://ui.adsabs.harvard.edu/abs/2019ApJ...874..153B} {874, 153}

\bibitem[\protect\citeauthoryear{{Callingham} et~al.,}{{Callingham}
  et~al.}{2017}]{Callingham}
{Callingham} J.~R.,  et~al., 2017, \mn@doi [\apj]
  {10.3847/1538-4357/836/2/174}, \href
  {https://ui.adsabs.harvard.edu/abs/2017ApJ...836..174C} {836, 174}

\bibitem[\protect\citeauthoryear{{Callister}, {Anderson}  \&
  {Hallinan}}{{Callister} et~al.}{2019}]{2019APS..APRQ16006C}
{Callister} T.,  {Anderson} M.,   {Hallinan} G.,  2019, in APS April Meeting
  Abstracts. p. Q16.006

\bibitem[\protect\citeauthoryear{{Carvalho} et~al.,}{{Carvalho}
  et~al.}{2019}]{2019ICRC...36..211C}
{Carvalho} W.~R.,  et~al., 2019, in 36th International Cosmic Ray Conference
  (ICRC2019). p.~211

\bibitem[\protect\citeauthoryear{{Chatterjee} \& {Bharadwaj}}{{Chatterjee} \&
  {Bharadwaj}}{2019}]{2019MNRAS.483.2269C}
{Chatterjee} S.,  {Bharadwaj} S.,  2019, \mn@doi [\mnras]
  {10.1093/mnras/sty3242}, \href
  {http://adsabs.harvard.edu/abs/2019MNRAS.483.2269C} {483, 2269}

\bibitem[\protect\citeauthoryear{{Chhabra}, {Gary}, {Hallinan}, {Anderson}  \&
  {Chen}}{{Chhabra} et~al.}{2019}]{2019AGUFMSH21B..03C}
{Chhabra} S.,  {Gary} D.~E.,  {Hallinan} G.,  {Anderson} M.,   {Chen} B.,
  2019, in AGU Fall Meeting Abstracts. pp SH21B--03

\bibitem[\protect\citeauthoryear{{Cohen}, {Lane}, {Cotton}, {Kassim}, {Lazio},
  {Perley}, {Condon}  \& {Erickson}}{{Cohen}
  et~al.}{2007}]{2007AJ....134.1245C}
{Cohen} A.~S.,  {Lane} W.~M.,  {Cotton} W.~D.,  {Kassim} N.~E.,  {Lazio}
  T.~J.~W.,  {Perley} R.~A.,  {Condon} J.~J.,   {Erickson} W.~C.,  2007,
  \mn@doi [\aj] {10.1086/520719}, \href
  {https://ui.adsabs.harvard.edu/abs/2007AJ....134.1245C} {134, 1245}

\bibitem[\protect\citeauthoryear{{Cohen}, {Fialkov}, {Barkana}  \&
  {Lotem}}{{Cohen} et~al.}{2017}]{2017MNRAS.472.1915C}
{Cohen} A.,  {Fialkov} A.,  {Barkana} R.,   {Lotem} M.,  2017, \mn@doi [\mnras]
  {10.1093/mnras/stx2065}, \href
  {http://adsabs.harvard.edu/abs/2017MNRAS.472.1915C} {472, 1915}

\bibitem[\protect\citeauthoryear{{Cohen}, {Fialkov}  \& {Barkana}}{{Cohen}
  et~al.}{2018}]{2018MNRAS.478.2193C}
{Cohen} A.,  {Fialkov} A.,   {Barkana} R.,  2018, \mn@doi [\mnras]
  {10.1093/mnras/sty1094}, \href
  {https://ui.adsabs.harvard.edu/abs/2018MNRAS.478.2193C} {478, 2193}

\bibitem[\protect\citeauthoryear{{Cohen}, {Fialkov}, {Barkana}  \&
  {Monsalve}}{{Cohen} et~al.}{2020}]{2019arXiv191006274C}
{Cohen} A.,  {Fialkov} A.,  {Barkana} R.,   {Monsalve} R.~A.,  2020, \mn@doi
  [\mnras] {10.1093/mnras/staa1530}, \href
  {https://ui.adsabs.harvard.edu/abs/2020MNRAS.495.4845C} {495, 4845}

\bibitem[\protect\citeauthoryear{{DeBoer} et~al.,}{{DeBoer}
  et~al.}{2017}]{deboer17}
{DeBoer} D.~R.,  et~al., 2017, \mn@doi [\pasp]
  {10.1088/1538-3873/129/974/045001}, \href
  {http://adsabs.harvard.edu/abs/2017PASP..129d5001D} {129, 045001}

\bibitem[\protect\citeauthoryear{{Dowell}}{{Dowell}}{2011}]{DowellMemo}
{Dowell} J.,  2011, {Parametric Model for the LWA-1 Dipole Response as a
  Function of Frequency},
  \url{https://www.faculty.ece.vt.edu/swe/lwa/memo/lwa0178.pdf}

\bibitem[\protect\citeauthoryear{{Dowell}, {Taylor}, {Schinzel}, {Kassim}  \&
  {Stovall}}{{Dowell} et~al.}{2017}]{2017MNRAS.469.4537D}
{Dowell} J.,  {Taylor} G.~B.,  {Schinzel} F.~K.,  {Kassim} N.~E.,   {Stovall}
  K.,  2017, \mn@doi [\mnras] {10.1093/mnras/stx1136}, \href
  {http://adsabs.harvard.edu/abs/2017MNRAS.469.4537D} {469, 4537}

\bibitem[\protect\citeauthoryear{{Eastwood} \& {Hallinan}}{{Eastwood} \&
  {Hallinan}}{2018}]{2018IAUS..333..110E}
{Eastwood} M.~W.,  {Hallinan} G.,  2018, in {Jeli{\'c}} V.,  {van der Hulst}
  T.,  eds,  IAU Symposium Vol. 333, Peering towards Cosmic Dawn. pp 110--113,
  \mn@doi{10.1017/S1743921317011231}

\bibitem[\protect\citeauthoryear{{Eastwood} et~al.,}{{Eastwood}
  et~al.}{2018}]{2018AJ....156...32E}
{Eastwood} M.~W.,  et~al., 2018, \mn@doi [\aj] {10.3847/1538-3881/aac721},
  \href {https://ui.adsabs.harvard.edu/abs/2018AJ....156...32E} {156, 32}

\bibitem[\protect\citeauthoryear{{Eastwood} et~al.,}{{Eastwood}
  et~al.}{2019}]{2019AJ....158...84E}
{Eastwood} M.~W.,  et~al., 2019, \mn@doi [\aj] {10.3847/1538-3881/ab2629},
  \href {https://ui.adsabs.harvard.edu/abs/2019AJ....158...84E} {158, 84}

\bibitem[\protect\citeauthoryear{{Ellingson}}{{Ellingson}}{2011}]{2011ITAP...59.1855E}
{Ellingson} S.~W.,  2011, \mn@doi [IEEE Transactions on Antennas and
  Propagation] {10.1109/TAP.2011.2122230}, \href
  {https://ui.adsabs.harvard.edu/abs/2011ITAP...59.1855E} {59, 1855}

\bibitem[\protect\citeauthoryear{{Ewall-Wice} et~al.,}{{Ewall-Wice}
  et~al.}{2016}]{EwallWice:2016}
{Ewall-Wice} A.,  et~al., 2016, \mn@doi [\mnras] {10.1093/mnras/stw1022}, \href
  {https://ui.adsabs.harvard.edu/abs/2016MNRAS.460.4320E} {460, 4320}

\bibitem[\protect\citeauthoryear{{Fan} et~al.,}{{Fan} et~al.}{2006}]{Fan:2006}
{Fan} X.,  et~al., 2006, \mn@doi [\aj] {10.1086/500296}, \href
  {https://ui.adsabs.harvard.edu/abs/2006AJ....131.1203F} {131, 1203}

\bibitem[\protect\citeauthoryear{{Feng} \& {Holder}}{{Feng} \&
  {Holder}}{2018}]{Feng:2018}
{Feng} C.,  {Holder} G.,  2018, \mn@doi [The Astrophysical Journal]
  {10.3847/2041-8213/aac0fe}, \href
  {https://ui.adsabs.harvard.edu/abs/2018ApJ...858L..17F} {858, L17}

\bibitem[\protect\citeauthoryear{{Fialkov} \& {Barkana}}{{Fialkov} \&
  {Barkana}}{2019}]{2019arXiv190202438F}
{Fialkov} A.,  {Barkana} R.,  2019, \mn@doi [\mnras] {10.1093/mnras/stz873},
  \href {https://ui.adsabs.harvard.edu/abs/2019MNRAS.486.1763F} {486, 1763}

\bibitem[\protect\citeauthoryear{{Fialkov}, {Barkana}, {Visbal},
  {Tseliakhovich}  \& {Hirata}}{{Fialkov} et~al.}{2013}]{Fialkov:2013}
{Fialkov} A.,  {Barkana} R.,  {Visbal} E.,  {Tseliakhovich} D.,   {Hirata}
  C.~M.,  2013, \mn@doi [\mnras] {10.1093/mnras/stt650}, \href
  {https://ui.adsabs.harvard.edu/abs/2013MNRAS.432.2909F} {432, 2909}

\bibitem[\protect\citeauthoryear{{Fialkov}, {Barkana}  \& {Cohen}}{{Fialkov}
  et~al.}{2018}]{Fialkov:2018}
{Fialkov} A.,  {Barkana} R.,   {Cohen} A.,  2018, \mn@doi [\prl]
  {10.1103/PhysRevLett.121.011101}, \href
  {https://ui.adsabs.harvard.edu/abs/2018PhRvL.121a1101F} {121, 011101}

\bibitem[\protect\citeauthoryear{{Field}}{{Field}}{1958}]{Field:1958}
{Field} G.~B.,  1958, \mn@doi [Proceedings of the IRE]
  {10.1109/JRPROC.1958.286741}, \href
  {https://ui.adsabs.harvard.edu/abs/1958PIRE...46..240F} {46, 240}

\bibitem[\protect\citeauthoryear{{Furlanetto}, {Oh}  \& {Briggs}}{{Furlanetto}
  et~al.}{2006}]{2006PhR...433..181F}
{Furlanetto} S.~R.,  {Oh} S.~P.,   {Briggs} F.~H.,  2006, \mn@doi [\physrep]
  {10.1016/j.physrep.2006.08.002}, \href
  {http://adsabs.harvard.edu/abs/2006PhR...433..181F} {433, 181}

\bibitem[\protect\citeauthoryear{{Gehlot} et~al.,}{{Gehlot}
  et~al.}{2019}]{Gehlot:2019}
{Gehlot} B.~K.,  et~al., 2019, \mn@doi [\mnras] {10.1093/mnras/stz1937}, \href
  {https://ui.adsabs.harvard.edu/abs/2019MNRAS.488.4271G} {488, 4271}

\bibitem[\protect\citeauthoryear{{Gehlot} et~al.,}{{Gehlot}
  et~al.}{2020}]{2020MNRAS.499.4158G}
{Gehlot} B.~K.,  et~al., 2020, \mn@doi [\mnras] {10.1093/mnras/staa3093}, \href
  {https://ui.adsabs.harvard.edu/abs/2020MNRAS.499.4158G} {499, 4158}

\bibitem[\protect\citeauthoryear{{Greig}, {Mesinger}, {Haiman}  \&
  {Simcoe}}{{Greig} et~al.}{2017}]{Greig:2017}
{Greig} B.,  {Mesinger} A.,  {Haiman} Z.,   {Simcoe} R.~A.,  2017, \mn@doi
  [\mnras] {10.1093/mnras/stw3351}, \href
  {https://ui.adsabs.harvard.edu/abs/2017MNRAS.466.4239G} {466, 4239}

\bibitem[\protect\citeauthoryear{{Greig}, {Mesinger}  \& {Ba{\~n}ados}}{{Greig}
  et~al.}{2019}]{Greig:2019}
{Greig} B.,  {Mesinger} A.,   {Ba{\~n}ados} E.,  2019, \mn@doi [\mnras]
  {10.1093/mnras/stz230}, \href
  {https://ui.adsabs.harvard.edu/abs/2019MNRAS.484.5094G} {484, 5094}

\bibitem[\protect\citeauthoryear{{Hallinan} \& {Anderson}}{{Hallinan} \&
  {Anderson}}{2017}]{2017reph.conf40101H}
{Hallinan} G.,  {Anderson} M.~M.,  2017, in Radio Exploration of Planetary
  Habitability (AASTCS5). p. 401.01

\bibitem[\protect\citeauthoryear{{Hallinan} et~al.,}{{Hallinan}
  et~al.}{2015}]{2015AAS...22532801H}
{Hallinan} G.,  et~al., 2015, in American Astronomical Society Meeting
  Abstracts \#225. p. 328.01

\bibitem[\protect\citeauthoryear{{Hills}, {Kulkarni}, {Meerburg}  \&
  {Puchwein}}{{Hills} et~al.}{2018}]{Hills:2018}
{Hills} R.,  {Kulkarni} G.,  {Meerburg} P.~D.,   {Puchwein} E.,  2018, \mn@doi
  [\nat] {10.1038/s41586-018-0796-5}, \href
  {https://ui.adsabs.harvard.edu/abs/2018Natur.564E..32H} {564, E32}

\bibitem[\protect\citeauthoryear{{H{\"o}gbom}}{{H{\"o}gbom}}{1974}]{1974A&AS...15..417H}
{H{\"o}gbom} J.~A.,  1974, \aaps, \href
  {https://ui.adsabs.harvard.edu/abs/1974A&AS...15..417H} {15, 417}

\bibitem[\protect\citeauthoryear{{Hogg}}{{Hogg}}{1999}]{1999astro.ph..5116H}
{Hogg} D.~W.,  1999, arXiv e-prints, \href
  {https://ui.adsabs.harvard.edu/abs/1999astro.ph..5116H} {astro-ph/9905116v4}

\bibitem[\protect\citeauthoryear{{Kern} et~al.,}{{Kern}
  et~al.}{2020a}]{2020ApJ...888...70K}
{Kern} N.~S.,  et~al., 2020a, \mn@doi [\apj] {10.3847/1538-4357/ab5e8a}, \href
  {https://ui.adsabs.harvard.edu/abs/2020ApJ...888...70K} {888, 70}

\bibitem[\protect\citeauthoryear{{Kern} et~al.,}{{Kern}
  et~al.}{2020b}]{2020ApJ...890..122K}
{Kern} N.~S.,  et~al., 2020b, \mn@doi [\apj] {10.3847/1538-4357/ab67bc}, \href
  {https://ui.adsabs.harvard.edu/abs/2020ApJ...890..122K} {890, 122}

\bibitem[\protect\citeauthoryear{{Kocz} et~al.,}{{Kocz}
  et~al.}{2015}]{2015JAI.....450003K}
{Kocz} J.,  et~al., 2015, \mn@doi [Journal of Astronomical Instrumentation]
  {10.1142/S2251171715500038}, \href
  {http://adsabs.harvard.edu/abs/2015JAI.....450003K} {4, 1550003}

\bibitem[\protect\citeauthoryear{{Kolopanis} et~al.,}{{Kolopanis}
  et~al.}{2019}]{Kolopanis:2019}
{Kolopanis} M.,  et~al., 2019, \mn@doi [\apj] {10.3847/1538-4357/ab3e3a}, \href
  {https://ui.adsabs.harvard.edu/abs/2019ApJ...883..133K} {883, 133}

\bibitem[\protect\citeauthoryear{{Koopmans} et~al.,}{{Koopmans}
  et~al.}{2015}]{Koopmans:2015}
{Koopmans} L.,  et~al., 2015, in Advancing Astrophysics with the Square
  Kilometre Array (AASKA14). p.~1 (\mn@eprint {arXiv} {1505.07568})

\bibitem[\protect\citeauthoryear{{Lanman}, {Pober}, {Kern}, {Acedo}, {DeBoer}
  \& {Fagnoni}}{{Lanman} et~al.}{2020}]{2020MNRAS.tmp.1168L}
{Lanman} A.~E.,  {Pober} J.~C.,  {Kern} N.~S.,  {Acedo} E. d.~L.,  {DeBoer}
  D.~R.,   {Fagnoni} N.,  2020, \mn@doi [\mnras] {10.1093/mnras/staa987}, \href
  {https://ui.adsabs.harvard.edu/abs/2020MNRAS.tmp.1168L} {}

\bibitem[\protect\citeauthoryear{{Li} et~al.,}{{Li} et~al.}{2019}]{Li:2019}
{Li} W.,  et~al., 2019, \mn@doi [\apj] {10.3847/1538-4357/ab55e4}, \href
  {https://ui.adsabs.harvard.edu/abs/2019ApJ...887..141L} {887, 141}

\bibitem[\protect\citeauthoryear{{Liu}, {Parsons}  \& {Trott}}{{Liu}
  et~al.}{2014}]{2014PhRvD..90b3018L}
{Liu} A.,  {Parsons} A.~R.,   {Trott} C.~M.,  2014, \mn@doi [\prd]
  {10.1103/PhysRevD.90.023018}, \href
  {http://adsabs.harvard.edu/abs/2014PhRvD..90b3018L} {90, 023018}

\bibitem[\protect\citeauthoryear{{Madau}, {Meiksin}  \& {Rees}}{{Madau}
  et~al.}{1997}]{Madau:1997}
{Madau} P.,  {Meiksin} A.,   {Rees} M.~J.,  1997, \mn@doi [\apj]
  {10.1086/303549}, \href
  {https://ui.adsabs.harvard.edu/abs/1997ApJ...475..429M} {475, 429}

\bibitem[\protect\citeauthoryear{{Mason}, {Treu}, {Dijkstra}, {Mesinger},
  {Trenti}, {Pentericci}, {de Barros}  \& {Vanzella}}{{Mason}
  et~al.}{2018}]{Mason:2018}
{Mason} C.~A.,  {Treu} T.,  {Dijkstra} M.,  {Mesinger} A.,  {Trenti} M.,
  {Pentericci} L.,  {de Barros} S.,   {Vanzella} E.,  2018, \mn@doi [\apj]
  {10.3847/1538-4357/aab0a7}, \href
  {https://ui.adsabs.harvard.edu/abs/2018ApJ...856....2M} {856, 2}

\bibitem[\protect\citeauthoryear{{Mertens} et~al.,}{{Mertens}
  et~al.}{2020}]{LOFAR-EoR:2020}
{Mertens} F.~G.,  et~al., 2020, \mn@doi [\mnras] {10.1093/mnras/staa327}, \href
  {https://ui.adsabs.harvard.edu/abs/2020MNRAS.493.1662M} {493, 1662}

\bibitem[\protect\citeauthoryear{{Mitchell}, {Greenhill}, {Wayth}, {Sault},
  {Lonsdale}, {Cappallo}, {Morales}  \& {Ord}}{{Mitchell}
  et~al.}{2008}]{2008ISTSP...2..707M}
{Mitchell} D.~A.,  {Greenhill} L.~J.,  {Wayth} R.~B.,  {Sault} R.~J.,
  {Lonsdale} C.~J.,  {Cappallo} R.~J.,  {Morales} M.~F.,   {Ord} S.~M.,  2008,
  \mn@doi [IEEE Journal of Selected Topics in Signal Processing]
  {10.1109/JSTSP.2008.2005327}, \href
  {http://adsabs.harvard.edu/abs/2008ISTSP...2..707M} {2, 707}

\bibitem[\protect\citeauthoryear{{Mitchell} et~al.,}{{Mitchell}
  et~al.}{2010}]{2010rfim.workE..16M}
{Mitchell} D.,  et~al., 2010, in RFI Mitigation Workshop. p.~16 (\mn@eprint
  {arXiv} {1008.2551})

\bibitem[\protect\citeauthoryear{{Monroe} et~al.,}{{Monroe}
  et~al.}{2020}]{2020NIMPA.95363086M}
{Monroe} R.,  et~al., 2020, \mn@doi [Nuclear Instruments and Methods in Physics
  Research A] {10.1016/j.nima.2019.163086}, \href
  {https://ui.adsabs.harvard.edu/abs/2020NIMPA.95363086M} {953, 163086}

\bibitem[\protect\citeauthoryear{{Monsalve}, {Rogers}, {Bowman}  \&
  {Mozdzen}}{{Monsalve} et~al.}{2017}]{Monsalve:2017}
{Monsalve} R.~A.,  {Rogers} A. E.~E.,  {Bowman} J.~D.,   {Mozdzen} T.~J.,
  2017, \mn@doi [\apj] {10.3847/1538-4357/aa88d1}, \href
  {https://ui.adsabs.harvard.edu/abs/2017ApJ...847...64M} {847, 64}

\bibitem[\protect\citeauthoryear{{Morales}, {Beardsley}, {Pober}, {Barry},
  {Hazelton}, {Jacobs}  \& {Sullivan}}{{Morales}
  et~al.}{2019}]{2019MNRAS.483.2207M}
{Morales} M.~F.,  {Beardsley} A.,  {Pober} J.,  {Barry} N.,  {Hazelton} B.,
  {Jacobs} D.,   {Sullivan} I.,  2019, \mn@doi [\mnras]
  {10.1093/mnras/sty2844}, \href
  {https://ui.adsabs.harvard.edu/abs/2019MNRAS.483.2207M} {483, 2207}

\bibitem[\protect\citeauthoryear{{Neben} et~al.,}{{Neben}
  et~al.}{2015}]{2015RaSc...50..614N}
{Neben} A.~R.,  et~al., 2015, \mn@doi [Radio Science] {10.1002/2015RS005678},
  \href {https://ui.adsabs.harvard.edu/abs/2015RaSc...50..614N} {50, 614}

\bibitem[\protect\citeauthoryear{{Offringa}, {van de Gronde}  \&
  {Roerdink}}{{Offringa} et~al.}{2012}]{2012A&A...539A..95O}
{Offringa} A.~R.,  {van de Gronde} J.~J.,   {Roerdink} J.~B.~T.~M.,  2012,
  \mn@doi [\aap] {10.1051/0004-6361/201118497}, \href
  {https://ui.adsabs.harvard.edu/abs/2012A&A...539A..95O} {539, A95}

\bibitem[\protect\citeauthoryear{{Paciga} et~al.,}{{Paciga}
  et~al.}{2013}]{Paciga:2013}
{Paciga} G.,  et~al., 2013, \mn@doi [\mnras] {10.1093/mnras/stt753}, \href
  {https://ui.adsabs.harvard.edu/abs/2013MNRAS.433..639P} {433, 639}

\bibitem[\protect\citeauthoryear{{Parsons}}{{Parsons}}{2017}]{ParsonsMemo}
{Parsons} A.,  2017, {Power Spectrum Normalizations for HERA},
  \url{http://reionization.org/wp-content/uploads/2013/03/Power_Spectrum_Normalizations_for_HERA.pdf}

\bibitem[\protect\citeauthoryear{{Parsons}, {Pober}, {McQuinn}, {Jacobs}  \&
  {Aguirre}}{{Parsons} et~al.}{2012a}]{2012ApJ...753...81P}
{Parsons} A.,  {Pober} J.,  {McQuinn} M.,  {Jacobs} D.,   {Aguirre} J.,  2012a,
  \mn@doi [\apj] {10.1088/0004-637X/753/1/81}, \href
  {http://adsabs.harvard.edu/abs/2012ApJ...753...81P} {753, 81}

\bibitem[\protect\citeauthoryear{{Parsons}, {Pober}, {Aguirre}, {Carilli},
  {Jacobs}  \& {Moore}}{{Parsons} et~al.}{2012b}]{2012ApJ...756..165P}
{Parsons} A.~R.,  {Pober} J.~C.,  {Aguirre} J.~E.,  {Carilli} C.~L.,  {Jacobs}
  D.~C.,   {Moore} D.~F.,  2012b, \mn@doi [\apj] {10.1088/0004-637X/756/2/165},
  \href {http://adsabs.harvard.edu/abs/2012ApJ...756..165P} {756, 165}

\bibitem[\protect\citeauthoryear{{Parsons} et~al.,}{{Parsons}
  et~al.}{2014}]{2014ApJ...788..106P}
{Parsons} A.~R.,  et~al., 2014, \mn@doi [\apj] {10.1088/0004-637X/788/2/106},
  \href {https://ui.adsabs.harvard.edu/abs/2014ApJ...788..106P} {788, 106}

\bibitem[\protect\citeauthoryear{{Parsons}, {Liu}, {Ali}  \& {Cheng}}{{Parsons}
  et~al.}{2016}]{2016ApJ...820...51P}
{Parsons} A.~R.,  {Liu} A.,  {Ali} Z.~S.,   {Cheng} C.,  2016, \mn@doi [\apj]
  {10.3847/0004-637X/820/1/51}, \href
  {http://adsabs.harvard.edu/abs/2016ApJ...820...51P} {820, 51}

\bibitem[\protect\citeauthoryear{{Paul} et~al.,}{{Paul}
  et~al.}{2014}]{2014ApJ...793...28P}
{Paul} S.,  et~al., 2014, \mn@doi [\apj] {10.1088/0004-637X/793/1/28}, \href
  {https://ui.adsabs.harvard.edu/abs/2014ApJ...793...28P} {793, 28}

\bibitem[\protect\citeauthoryear{{Paul} et~al.,}{{Paul}
  et~al.}{2016}]{2016ApJ...833..213P}
{Paul} S.,  et~al., 2016, \mn@doi [\apj] {10.3847/1538-4357/833/2/213}, \href
  {https://ui.adsabs.harvard.edu/abs/2016ApJ...833..213P} {833, 213}

\bibitem[\protect\citeauthoryear{{Planck Collaboration} et~al.,}{{Planck
  Collaboration} et~al.}{2018}]{Planck:2018}
{Planck Collaboration} et~al., 2018, arXiv e-prints, \href
  {https://ui.adsabs.harvard.edu/abs/2018arXiv180706209P} {p. arXiv:1807.06209}

\bibitem[\protect\citeauthoryear{{Pober} et~al.,}{{Pober}
  et~al.}{2013}]{2013ApJ...768L..36P}
{Pober} J.~C.,  et~al., 2013, \mn@doi [\apjl] {10.1088/2041-8205/768/2/L36},
  \href {http://adsabs.harvard.edu/abs/2013ApJ...768L..36P} {768, L36}

\bibitem[\protect\citeauthoryear{{Price} et~al.,}{{Price}
  et~al.}{2018}]{2018MNRAS.478.4193P}
{Price} D.~C.,  et~al., 2018, \mn@doi [\mnras] {10.1093/mnras/sty1244}, \href
  {https://ui.adsabs.harvard.edu/abs/2018MNRAS.478.4193P} {478, 4193}

\bibitem[\protect\citeauthoryear{{Reis}, {Fialkov}  \& {Barkana}}{{Reis}
  et~al.}{2020}]{2020MNRAS.499.5993R}
{Reis} I.,  {Fialkov} A.,   {Barkana} R.,  2020, \mn@doi [\mnras]
  {10.1093/mnras/staa3091}, \href
  {https://ui.adsabs.harvard.edu/abs/2020MNRAS.499.5993R} {499, 5993}

\bibitem[\protect\citeauthoryear{{Romero-Wolf} et~al.,}{{Romero-Wolf}
  et~al.}{2019}]{2019ICRC...36..405R}
{Romero-Wolf} A.,  et~al., 2019, in 36th International Cosmic Ray Conference
  (ICRC2019). p.~405

\bibitem[\protect\citeauthoryear{{Shaw}, {Sigurdson}, {Pen}, {Stebbins}  \&
  {Sitwell}}{{Shaw} et~al.}{2014}]{Shaw:2014}
{Shaw} J.~R.,  {Sigurdson} K.,  {Pen} U.-L.,  {Stebbins} A.,   {Sitwell} M.,
  2014, \mn@doi [\apj] {10.1088/0004-637X/781/2/57}, \href
  {https://ui.adsabs.harvard.edu/abs/2014ApJ...781...57S} {781, 57}

\bibitem[\protect\citeauthoryear{{Shume}, {Hallinan}, {Anderson}, {Monroe}  \&
  {Eastwood}}{{Shume} et~al.}{2017}]{2017AGUFMSA21A2499S}
{Shume} E.~B.,  {Hallinan} G.,  {Anderson} M.,  {Monroe} R.,   {Eastwood} M.,
  2017, in AGU Fall Meeting Abstracts. pp SA21A--2499

\bibitem[\protect\citeauthoryear{{Sims} \& {Pober}}{{Sims} \&
  {Pober}}{2019}]{Sims:2019}
{Sims} P.~H.,  {Pober} J.~C.,  2019, arXiv e-prints, \href
  {https://ui.adsabs.harvard.edu/abs/2019arXiv191003165S} {p. arXiv:1910.03165}

\bibitem[\protect\citeauthoryear{{Singh} \& {Subrahmanyan}}{{Singh} \&
  {Subrahmanyan}}{2019}]{Singh:2019}
{Singh} S.,  {Subrahmanyan} R.,  2019, \mn@doi [\apj]
  {10.3847/1538-4357/ab2879}, \href
  {https://ui.adsabs.harvard.edu/abs/2019ApJ...880...26S} {880, 26}

\bibitem[\protect\citeauthoryear{{Singh} et~al.,}{{Singh}
  et~al.}{2017}]{Singh:2017}
{Singh} S.,  et~al., 2017, \mn@doi [\apjl] {10.3847/2041-8213/aa831b}, \href
  {https://ui.adsabs.harvard.edu/abs/2017ApJ...845L..12S} {845, L12}

\bibitem[\protect\citeauthoryear{{Spinelli}, {Bernardi}  \&
  {Santos}}{{Spinelli} et~al.}{2019}]{Spinelli:2019}
{Spinelli} M.,  {Bernardi} G.,   {Santos} M.~G.,  2019, \mn@doi [\mnras]
  {10.1093/mnras/stz2425}, \href
  {https://ui.adsabs.harvard.edu/abs/2019MNRAS.489.4007S} {489, 4007}

\bibitem[\protect\citeauthoryear{{Spinelli}, {Bernardi}, {Garsden}, {Greehill},
  {Fialkov}, {Dowell}  \& {Price}}{{Spinelli}
  et~al.}{2020}]{2020arXiv201103994S}
{Spinelli} M.,  {Bernardi} G.,  {Garsden} H.,  {Greehill} L.~J.,  {Fialkov} A.,
   {Dowell} J.,   {Price} D.~C.,  2020, arXiv e-prints, \href
  {https://ui.adsabs.harvard.edu/abs/2020arXiv201103994S} {p. arXiv:2011.03994}

\bibitem[\protect\citeauthoryear{{Taylor} et~al.,}{{Taylor}
  et~al.}{2012}]{2012JAI.....150004T}
{Taylor} G.~B.,  et~al., 2012, \mn@doi [Journal of Astronomical
  Instrumentation] {10.1142/S2251171712500043}, \href
  {http://adsabs.harvard.edu/abs/2012JAI.....150004T} {1, 1250004}

\bibitem[\protect\citeauthoryear{{Thompson}, {Moran}  \& {Swenson}}{{Thompson}
  et~al.}{2017}]{2017isra.book.....T}
{Thompson} A.~R.,  {Moran} J.~M.,   {Swenson} George~W. J.,  2017,
  {Interferometry and Synthesis in Radio Astronomy, 3rd Edition}.
Springer, \mn@doi{10.1007/978-3-319-44431-4}

\bibitem[\protect\citeauthoryear{{Thyagarajan} et~al.,}{{Thyagarajan}
  et~al.}{2015a}]{2015ApJ...804...14T}
{Thyagarajan} N.,  et~al., 2015a, \mn@doi [\apj] {10.1088/0004-637X/804/1/14},
  \href {http://adsabs.harvard.edu/abs/2015ApJ...804...14T} {804, 14}

\bibitem[\protect\citeauthoryear{{Thyagarajan} et~al.,}{{Thyagarajan}
  et~al.}{2015b}]{2015ApJ...807L..28T}
{Thyagarajan} N.,  et~al., 2015b, \mn@doi [\apjl]
  {10.1088/2041-8205/807/2/L28}, \href
  {http://adsabs.harvard.edu/abs/2015ApJ...807L..28T} {807, L28}

\bibitem[\protect\citeauthoryear{{Trott} et~al.,}{{Trott}
  et~al.}{2020}]{Trott:2020}
{Trott} C.~M.,  et~al., 2020, \mn@doi [\mnras] {10.1093/mnras/staa414}, \href
  {https://ui.adsabs.harvard.edu/abs/2020MNRAS.493.4711T} {493, 4711}

\bibitem[\protect\citeauthoryear{{Ung}, {Sokolowski}, {Sutinjo}  \&
  {Davidson}}{{Ung} et~al.}{2020}]{2020arXiv200305116U}
{Ung} D. C.~X.,  {Sokolowski} M.,  {Sutinjo} A.~T.,   {Davidson} D.~B.,  2020,
  arXiv e-prints, \href {https://ui.adsabs.harvard.edu/abs/2020arXiv200305116U}
  {p. arXiv:2003.05116}

\bibitem[\protect\citeauthoryear{{{\"U}st{\"u}ner}, {Aydemir}, {G{\"u}le{\c
  c}}, {İlarslan}, {{\c C}elebi}  \& {Demirel}}{{{\"U}st{\"u}ner}
  et~al.}{2014}]{6929024}
{{\"U}st{\"u}ner} F.,  {Aydemir} E.,  {G{\"u}le{\c c}} E.,  {İlarslan} M.,
  {{\c C}elebi} M.,   {Demirel} E.,  2014, in 2014 XXXIth URSI General Assembly
  and Scientific Symposium (URSI GASS). pp~1--4

\bibitem[\protect\citeauthoryear{{Weinberger}, {Haehnelt}  \&
  {Kulkarni}}{{Weinberger} et~al.}{2019}]{Weinberger:2019}
{Weinberger} L.~H.,  {Haehnelt} M.~G.,   {Kulkarni} G.,  2019, \mn@doi [\mnras]
  {10.1093/mnras/stz481}, \href
  {https://ui.adsabs.harvard.edu/abs/2019MNRAS.485.1350W} {485, 1350}

\bibitem[\protect\citeauthoryear{{Wouthuysen}}{{Wouthuysen}}{1952}]{Wouthuysen:1952}
{Wouthuysen} S.~A.,  1952, \mn@doi [The Astronomical Journal] {10.1086/106661},
  \href {https://ui.adsabs.harvard.edu/abs/1952AJ.....57R..31W} {57, 31}

\bibitem[\protect\citeauthoryear{{de Gasperin} et~al.,}{{de Gasperin}
  et~al.}{2012}]{2012A&A...547A..56D}
{de Gasperin} F.,  et~al., 2012, \mn@doi [\aap] {10.1051/0004-6361/201220209},
  \href {https://ui.adsabs.harvard.edu/abs/2012A&A...547A..56D} {547, A56}

\bibitem[\protect\citeauthoryear{{de Oliveira-Costa}, {Tegmark}, {Gaensler},
  {Jonas}, {Landecker}  \& {Reich}}{{de Oliveira-Costa}
  et~al.}{2008}]{2008MNRAS.388..247D}
{de Oliveira-Costa} A.,  {Tegmark} M.,  {Gaensler} B.~M.,  {Jonas} J.,
  {Landecker} T.~L.,   {Reich} P.,  2008, \mn@doi [\mnras]
  {10.1111/j.1365-2966.2008.13376.x}, \href
  {https://ui.adsabs.harvard.edu/abs/2008MNRAS.388..247D} {388, 247}

\makeatother
\end{thebibliography}




\appendix

\setcounter{table}{0}
\renewcommand{\thetable}{A\arabic{table}}
\section{Regions used to obtain power values from power spectra}
\label{appendix_regions}
\begin{table*}
\centering
    \begin{tabular}{l | p{0.3\linewidth} |p{0.3\linewidth}}
    \hline
    Regions  & Purpose & Motivation\\
    \hline\hline
     \raisebox{-0.9\totalheight}{\includegraphics[clip,width=2in]{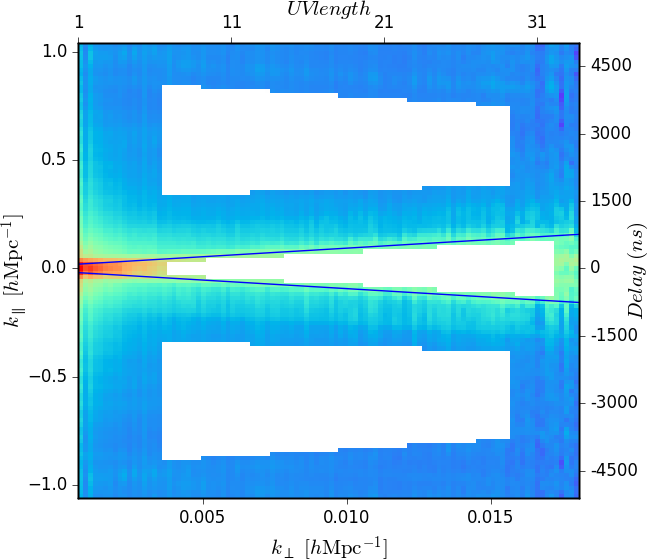}} &Applied only to power spectra generated from telescope observations (not simulations). These are for obtaining the mean value of $P(\mathbf{k})$ in the foreground wedge and Cosmic  Dawn window.  There are 2 Cosmic  Dawn window regions, and one foreground wedge region.
     &  The Cosmic  Dawn window  regions avoid the leakage from the wedge,
     and artifacts at large $k_\parallel$ (positive and negative). They
     cover a region where the power is fairly uniform, but still may include artifacts. The foreground wedge region  avoids the heightened power at low $k_\perp$.
    \\
     \hline
          \raisebox{-0.9\totalheight}{\includegraphics[clip,width=2in]{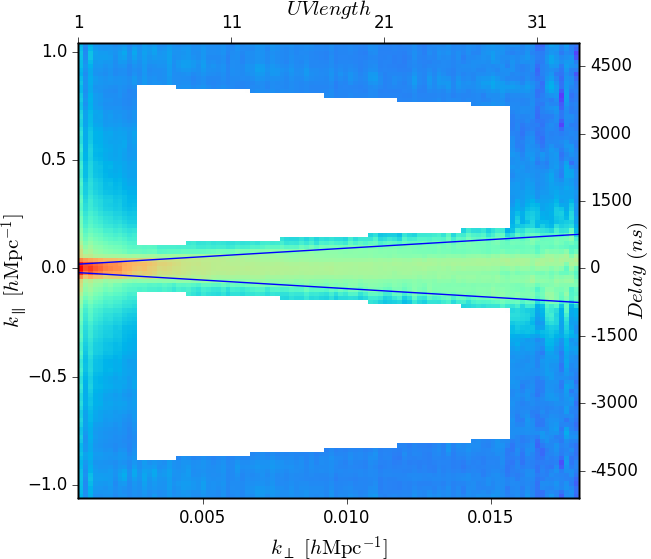}} &
     Applied only to power spectra generated from telescope observations (not simulations). When calculating $\Delta^2(k)$, use only power in these regions. 
     &  Similar to the previous Cosmic  Dawn window regions, but they include
     spillover  close to the horizon.
    \\
     \hline
          \raisebox{-0.9\totalheight}{\includegraphics[clip,width=2in]{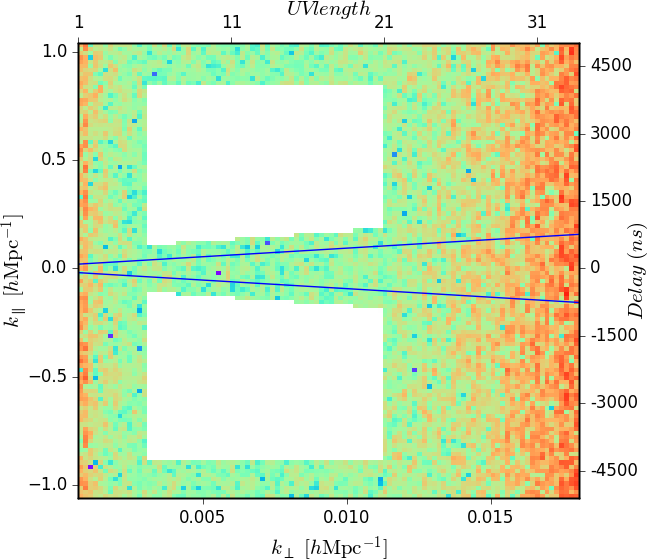}} &
     Applied only to  noise-only power spectra generated from simulated noise-only observations. For obtaining the mean value of $P(\mathbf{k})$ in the Cosmic  Dawn window.
     &  The regions are chosen to correspond to areas of the power
     spectrum where the noise is uniform and has its lowest value. The noise is not uniform over the power spectrum, there is a slight increase at $k_\parallel = 0$, and an increase at low and high $k_\perp$ because the density of baselines there is low.
    \\
     \hline
    \raisebox{-0.9\totalheight}{\includegraphics[clip,width=2in]{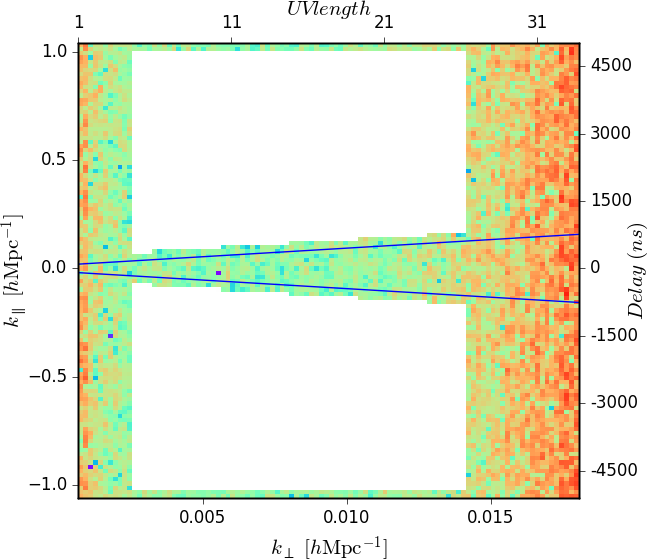}} &
     Applied only to  noise-only power spectra generated from simulated noise-only observations. When calculating $\Delta^2(k)$, use only values in these regions. They expand
     the regions shown in the previous row, so that  $\Delta^2(k)$ can be calculated for a wider range of $k$.
     &  Similar to the  Cosmic  Dawn window regions in row 2, but they extend close to the top/bottom edges of the power spectrum.
    \\
     \hline
    \end{tabular}
    \caption{Regions in the power spectrum used for obtaining power $P(\mathbf{k})$ and $\Delta^2(k)$}
    \label{regions}
\end{table*}
When extracting values from a power spectrum, for example to report statistics or generate $\Delta^2(k)$,  we use only certain regions. The regions avoid artifacts in the power spectrum and select areas appropriate for the required statistic; for example, statistics for the Cosmic  Dawn window avoid the foreground wedge and high power at low $k_\perp$. Table \ref{regions} shows the regions used for each type of data analysis; they are the area in white plotted on top of a power spectrum.   The first two rows in the table apply only to power spectra generated from observations, the next two rows apply only to power spectra generated from simulations of noise.

When a value for $P(\mathbf{k})$ is reported for a power spectrum, this is the average value of $P(\mathbf{k})$ within the region being discussed. To obtain $\Delta(k)^2$ values from a region, all the pixels within the region are assigned a $k$ value, converted to dimensionless power via Equation \ref{delta_eqn}, and binned (averaged) by $k$ using 16 logarithmic bins within the range of the obtained $k$ values.

\section{Selection of telescope observations used for power spectrum generation}
\label{selection_snapshots}

The selection criteria are simple and heuristic.  Observations for integrated power spectra are selected  first on the accuracy of their calibration, and then on the quality of the power spectrum that each observation produces. 
 
Each 9\,s observation is calibrated, and a calibration score assigned, based on how closely the calibrated observation reproduces the  flux densities of the calibration sources. A score of 1 indicates perfect calibration, and we select all observations with a score of 0.8 to 1.2. 

Each  observation selected by the previous step is then assigned a score obtained from the power spectrum that the observation generates. A high score indicates: (a) That the power outside the wedge is low compared to the power inside the wedge, and (b) that the power outside the horizon is flat, i.e. does not vary much but maintains a constant value. The scores are  obtained from  mean and RMS values of the power in regions in the power spectrum (Table \ref{regions}, row 1).

The 4 hours of observations selected for  power spectrum generation are the 4 hours of observations that scored the highest in the previous step.

\section{Time averaging limit for OVRO-LWA}
\label{integration_time_limit}
Coherent integration of  visibilities recorded at different  times may result in amplitude loss, due to sky sources changing position relative to interferometer fringes. This is referred to as time-smearing or time-averaging loss. When observing continuously, only visibilities recorded within a short time range $t_A$  may be coherently integrated.

Following Equations 16.5 and 16.6 in \cite{2017isra.book.....T}, the fractional loss in  amplitude, $\xi$, in integrated visibilities observed over some time range $T$, for an east-west oriented baseline, can be expressed as
\begin{equation}
    \xi = 1-\mathrm{sinc}(\pi \omega_\mathrm{e} \mathrm{cos} \delta [D_\mathrm{EW} \nu/c] T),
\end{equation}
where $D_\mathrm{EW}$ is the length of the baseline,  $\omega_\mathrm{e}$ is the  sidereal rate,
$\delta$ the observing declination of a source, and
$\nu$ the observing frequency.  For an interferometer containing many baselines, the most restrictive value for $T$ is found by setting $D_\mathrm{EW}$ equal to the maximum east-west baseline length, and setting $\delta$ to 0.
 
 For the OVRO-LWA telescope observations that were used for power spectrum generation, we allow that the fractional loss in amplitude can be no more than 0.01. Therefore we set $\xi = 0.01$,  $D_\mathrm{EW} = 200$\,m, $\delta = 0$, and   $\nu = 48.324$\,MHz,  and solve for $T$, giving   $t_A =  33$\,s. Observations whose times are separated by no more than 33\,s may be coherently combined. We  increase it to 36\,s so as to be a multiple of  the OVRO-LWA observation integration time (9\,s). The time limit for other frequencies can be similarly obtained.


\bsp	
\label{lastpage}
\end{document}